\documentclass[a4paper,times]{cjeart}
\usepackage{multirow}
\usepackage{makecell}
\usepackage{array}
\usepackage{amssymb}
\usepackage{hyperref}
\usepackage{amsthm} 
\usepackage{pifont}

\usepackage{booktabs}  
\usepackage{tabularx}  
\usepackage{ragged2e}  
\usepackage{arydshln}

\graphicspath{{figures/}}
\usepackage{algorithm,algorithmic}

\makeatletter
\let\myorg@bibitem\bibitem
\def\bibitem#1#2\par{%
	\@ifundefined{bibitem@#1}{%
		\myorg@bibitem{#1}#2\par
	}{%
		\begingroup
		\color{\csname bibitem@#1\endcsname}%
		\myorg@bibitem{#1}#2\par
		\endgroup
	}%
}
\makeatother

\journaltype{Review}

\title[Large Model Enabled Embodied Intelligence for 6G Integrated Perception, Communication, and Computation Network]{Large Model Enabled Embodied Intelligence for 6G Integrated Perception, Communication, and Computation Network}

\author{%
Zhuoran Li\affilnums{1,2}, 
Zhen Gao\affilnums{3,4,5,6},
Xinhua Liu\affilnums{1,2},
Zheng Wang\affilnums{7},
Xiaotian Zhou\affilnums{8,9},
Lei Liu\affilnums{10},
Yongpeng Wu\affilnums{11},
Wei Feng\affilnums{12},
Yongming Huang\affilnums{13,14}
}

\affiliation{%
\affilnum{1}School of Information and Electronics, Beijing Institute of Technology, Beijing 100081, China\\
\affilnum{2}School of Interdisciplinary Science, Beijing Institute of Technology, Beijing 100081, China\\
\affilnum{3}State Key Laboratory of CNS/ATM and the MIIT Key Laboratory of Complex-Field Intelligent Sensing, Beijing 100081, China\\
\affilnum{4}Beijing Institute of Technology (BIT), BIT, Zhuhai 519088, China\\
\affilnum{5}Advanced Technology Research Institute, BIT (Jinan), Jinan 250307, China\\
\affilnum{6}Yangtze Delta Region Academy, BIT (Jiaxing), Jiaxing 314019, China\\
\affilnum{7}School of School of Information Science and Engineering, Southeast University, Nanjing 210096, China\\
\affilnum{8}Institute of Intelligent Communication Technologies, Shandong University, Jinan 250061, China\\
\affilnum{9}Shandong Key Laboratory of Intelligent Communication and Sensing-Computing Integration Shandong University, Jinan 250061, China\\
\affilnum{10} Zhejiang Provincial Key Laboratory of Information Processing, Communication and Networking, College of Information Science and Electronic Engineering, Zhejiang University, Hangzhou 310007, China\\
\affilnum{11} Department of Electronic Engineering, Shanghai Jiao Tong University, Minhang 200240, China \\
\affilnum{12} Department of Electronic Engineering, State Key Laboratory of Space Network and Communications, Tsinghua University, Beijing 100084, China\\
\affilnum{13} National Mobile Communications Research Laboratory, School of Information Science and Engineering, Southeast University, Nanjing 210096, China\\
\affilnum{14} Purple Mountain Laboratories, Nanjing 211111, China
}
\corrauth{Zhen Gao}
\email{gaozhen16@bit.edu.cn}
\receivetime{Received x xx, xxxx}
\accepttime{Accepted x xx, xxx}
\publishtime{Published x xx, xxxx}
\abstract{%
The advent of sixth-generation (6G) places intelligence at the core of wireless architecture, fusing perception, communication, and computation into a single closed-loop.
This paper argues that large artificial intelligence models (LAMs) can endow base stations with perception, reasoning, and acting capabilities, thus transforming them into intelligent base station agents (IBSAs). 
We first review the historical evolution of BSs from single-functional analog infrastructure to distributed, software-defined, and finally LAM-empowered IBSA, highlighting the accompanying changes in architecture, hardware platforms, and deployment.
We then present an IBSA architecture that couples a perception-cognition-execution pipeline with cloud-edge-end collaboration and parameter-efficient adaptation.
Subsequently,we study two representative scenarios: (i) cooperative vehicle-road perception for autonomous driving, and (ii) ubiquitous base station support for low-altitude uncrewed aerial vehicle safety monitoring and response against unauthorized drones.
On this basis, we analyze key enabling technologies spanning LAM design and training, efficient edge-cloud inference, multi-modal perception and actuation, as well as trustworthy security and governance. 
We further propose a holistic evaluation framework and benchmark considerations that jointly cover communication performance, perception accuracy, decision-making reliability, safety, and energy efficiency.
Finally, we distill open challenges on benchmarks, continual adaptation, trustworthy decision-making, and standardization.
Together, this work positions LAM-enabled IBSAs as a practical path toward integrated perception, communication, and computation native, safety-critical 6G systems.
}

\keywords{6G, integrated perception, communication, and computation, intelligent base station agent (IBSA), large AI model (LAM), multimodal fusion.}

\begin{document}

\maketitle

\section{Introduction}
\label{sec1}

\subsection{Background and Motivation}
In the sixth-generation (6G) era, intelligence is no longer an afterthought but a native feature woven into every network layer. Unlike past generations, 6G elevates artificial intelligence (AI) from a tool to a fundamental infrastructure component, enabling networks to learn, adapt, and self‑optimize in real time~\cite{ref_NativeAI_3}. This AI‑native paradigm, often termed ``native AI" or ``AI‑native wireless systems", envisions embedding machine‑learning models directly into the radio interface and control planes to achieve ultra‑low latency, high reliability, and context‑aware decision‑making~\cite{ref_NativeAI_2}.
Recent work on artificial general intelligence (AGI)-native networks argues that simply scaling large language or generative models is insufficient; true network autonomy requires modules for perception, causal world modeling, and action planning to handle unforeseen scenarios with common-sense reasoning~\cite{ref_NativeAI_2}. In this vision, large language models (LLMs) translate operator intents into network objectives, while specialized foundation models predict environmental dynamics and proactively adjust resource allocation.
A quintessential application of native AI in 6G is integrated sensing and communications (ISAC), where sensing and payload transmission share spectrum, hardware, and intelligence pipelines~\cite{ref_Huang2024}. By co-designing radar-like perception with adaptive beamforming, ISAC exemplifies how native AI can unlock mutual performance gains, turning base stations (BSs) into perceptive agents that both see and serve the environment~\cite{ref_ISAC}.
\begin{table*}[!t]
	\centering
	\caption{Definition of perception, communication, and computation in the LAM-enabled IBSA paradigm.}
	\label{tab:pcc_definition}
	\renewcommand{\arraystretch}{1.4} 
	\small 
	\begin{tabularx}{\textwidth}{@{} p{2.5cm} >{\RaggedRight}X >{\RaggedRight}X @{}}
		\toprule
		\textbf{Function} & \textbf{Core Mechanism via LAM} & \textbf{Strategic Value for 6G} \\
		\midrule
		
		\textbf{Perception} 
		& \textbf{Semantic Alignment \& Grounding:} LAMs (e.g., grounded 3D-LLMs) fuse heterogeneous data (lidar, RGB, RF) into unified semantic tokens, translating raw 3D bounding boxes into actionable context.
		& Enables the BS to ``understand" the physical environment rather than just sensing signals, providing the foundation for closed-loop control in tasks like autonomous driving. \\
		\cmidrule(l){2-3}
		
		\textbf{Communication} 
		& \textbf{Semantic-Aware Resource Management:} The control plane utilizes LAMs to perform context-aware decision-making. Beamforming and resource allocation are driven by high-level semantic understanding of targets (e.g., pedestrians, vehicles) rather than solely by channel statistics.
		& Transforms the communication system from a signal-processing pipeline into a semantic-aware intelligent agent, significantly enhancing efficiency and reliability in dynamic scenarios. \\
		\cmidrule(l){2-3}
		
		\textbf{Computation} 
		& \textbf{Edge Intelligence \& Reasoning:} Deploying LAMs at the BS edge brings general-purpose cognitive capabilities closer to the user. It supports complex, latency-sensitive tasks such as trajectory prediction, interference coordination, and causal reasoning.
		& Acts as the ``glue" of the integrated perception, communication, and computation loop, enabling the BS to function as an autonomous agent that performs real-time planning and execution without reliance on distant cloud resources. \\
		
		\bottomrule
	\end{tabularx}
\end{table*}

\begin{table*}[!t]
	\centering
	\caption{Comparison of LAMs and traditional small models for the integrated perception, communication, and computation paradigm.}
	\label{tab:lam_vs_small}
	\renewcommand{\arraystretch}{1.3} 
	\begin{tabularx}{\textwidth}{@{} l >{\RaggedRight}X >{\RaggedRight}X >{\RaggedRight}X @{}}
		\toprule
		\textbf{Feature} & \textbf{Large AI Models} & \textbf{Traditional Small Models} & \textbf{Strategic Value} \\
		\midrule
		\textbf{Core Capability} 
		& General-purpose reasoning, high-level planning, multi-modal semantic alignment.
		& Domain-specific, feature extraction, optimization for specific tasks.
		& A unified cognitive and decision-making core.
		\\ \addlinespace
		\textbf{Generalization}
		& Excellent zero-shot/few-shot capabilities, adaptable to unstructured environments.
		& Heavily reliant on training sets, poor generalization.
		& Adapts to the dynamic, highly variable, and non-stationary real-world environment.
		\\ \addlinespace
		\textbf{Data Processing}
		& Dynamic, adaptive cross-modal fusion mechanisms.
		& Static feature concatenation, lacks adaptive modulation.
		& Enables real-time fusion and comprehension of high-dimensional, heterogeneous sensory data.
		\\
		\bottomrule
	\end{tabularx}
\end{table*}
\subsection{Necessity of Large Models}
The evolution of 6G networks necessitates a transition from static resource management to dynamic, intelligent adaptation. Foundational frameworks for ISAC, such as those comprehensively reviewed in~\cite{ref_ISAC}, have successfully established the theoretical limits and signal processing principles for dual-functional networks. However, these conventional approaches are predominantly model-driven and rule-based. They rely on explicit mathematical modeling of the channel and environment, which often becomes intractable or inaccurate in highly dynamic, non-stationary, and semantically complex real-world scenarios. While traditional AI has been deployed to address these model deficits and handle non-linear complexities, these traditional AI systems are often designed as highly modular and single-function ``silos," focusing on optimization within specific domains, such as pure image recognition, natural language processing, or network routing. This fragmented intelligent architecture exhibits inherent limitations when facing complex, dynamic, and cross-modal challenges in the real world. With the advancement of the AGI paradigm, the strategic objective of the scientific community has shifted toward building unified systems that feature integrated perception, communication, and computation\cite{ref_xf_JSAC,ref_cl_Network,ref_cl_JSAC}.

The integrated perception, communication, and computation paradigm aims to dismantle traditional functional boundaries, leveraging ubiquitous BS infrastructure to drive its transformation from a single-function communication node into an intelligent hub. This evolution represents not merely a functional aggregation but a fundamental paradigm shift enabled by large AI models (LAMs), which serve as the ideal engine to break down information barriers and empower intelligent BS agents (IBSAs). Under this framework, perception evolves from raw signal acquisition to semantic understanding, where models like grounded three-dimensional (3D)-LLMs align heterogeneous sensor data (e.g., lidar, radio-frequency (RF)) for closed-loop reasoning. Communication shifts towards semantic-aware resource management, utilizing these high-level semantic insights, rather than relying solely on channel statistics, to optimize resource allocation and beamforming. Finally, computation is pushed to the edge, where LAMs function as the reasoning core to execute complex planning and autonomous decision-making with low latency. This synergistic relationship, detailed in Table~\ref{tab:pcc_definition}, establishes the theoretical foundation for the IBSA architecture.
\begin{figure*}[!t]
	\centering	
	
	\includegraphics[width=7in]{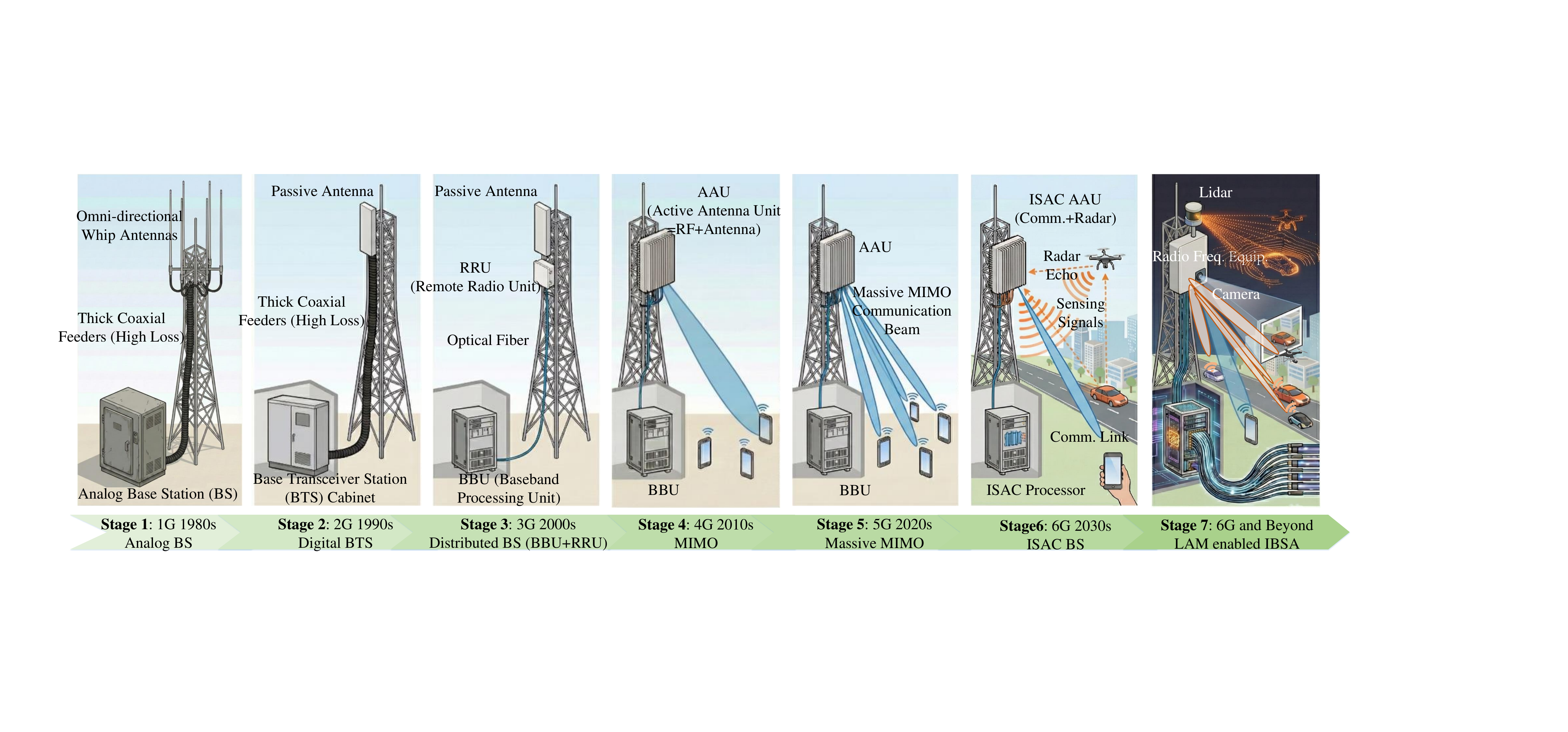}
	\caption{The physical evolution of the BS from 1G to 6G and beyond. The architecture has progressed from (Stage 1-2) centralized cabinets (analog BS, digital BTS) with high-loss coaxial feeders to (Stage 3) a distributed structure with a baseband processing unit (BBU) and remote radio unit (RRU) connected by optical fiber. This was followed by (Stage 4-5) the integration of RF and antenna into an active antenna unit (AAU) to support MIMO and massive MIMO. This integration path provides the foundation for (Stage 6) the 6G ISAC BS and (Stage 7) the LAM-enabled IBSA, which leverages multi-modal sensors like lidar and cameras for embodied intelligence.}
	\label{fig_BS}
\end{figure*}

Traditional small models are typically meticulously customized to solve specific benchmark tasks, relying on carefully tuned hyperparameters and algorithm selection~\cite{luhme2025proc}. In contrast, the success of LAMs stems from the scaling effect and the resulting emergent abilities. When model parameters and training data reach a certain threshold, the model's performance and functionality undergo a qualitative leap, a non-linear growth that traditional small models are unable to achieve~\cite{zhou2025perception}.
A systematic comparison highlighting these fundamental differences, specifically regarding core capabilities, generalization, and data processing mechanisms, is provided in Table~\ref{tab:lam_vs_small}.

LAMs pursue general-purpose solving strategies, which is reflected in the research trend toward seeking unified models to address diverse tasks. For example, some research is dedicated to building general-purpose, exemplary deep reinforcement learning (RL) algorithms that achieve competitive performance on a variety of RL benchmarks with a single set of hyperparameters~\cite{luhme2025proc}. The generality and zero-shot/few-shot learning capabilities of LAMs enable them to adapt to unstructured, highly variable real-world environments, which is fundamental to achieving general-purpose decision-making and cross-domain collaboration in integrated perception, communication, and computation.

In processing multimodal data, LAMs differ fundamentally from traditional methods. Traditional fusion mechanisms often rely on static feature integration, simply concatenating visual representations from different ``expert" encoders, and lack adaptive modulation capability~\cite{IoTJMFengZhiyong}. This rigid combination struggles to effectively address tasks requiring rich multi-scale understanding, such as fine-grained recognition, small object perception, or precise spatial localization. In contrast, multimodal LLM (M-LLMs) can leverage robust cross-modal transfer, in-context learning, and high-dimensional data processing capabilities to achieve dynamic and context-aware heterogeneous data fusion~\cite{IoTJMFengZhiyong}. This mechanism allows the system to flexibly weigh and interpret information from different modalities (such as visual, linguistic, and RF signals) based on current task and environmental demands, thereby enabling more advanced and adaptive decision-making than static integration and laying the technical foundation for real-time, complex decisions in integrated perception, communication, and computation.

\subsection{Structure of the Paper}
To begin with, Section~\ref{sec_evolution} traces the evolution of BSs from single-function infrastructure to LAM-enabled intelligent agents and provides a systematic comparison with existing next-generation architectures.
Subsequently, Section~\ref{sec_archi} introduces the overall architecture of embodied IBSA.
Next, Section~\ref{sec_scenario} elaborates on two key application scenarios of {\color{black}embodied} IBSA, namely cooperative vehicle-road autonomous driving (Section~\ref{sec_auto}) and ubiquitous {\color{black}embodied IBSA}-enabled low-altitude safety (Section~\ref{sec_LAE}).
Following that, Section~\ref{sec_enablingTech} discusses the enabling technologies that support the implementation of {\color{black}embodied} IBSA.
Subsequently, Section~\ref{sec_metrics} outlines the metrics and benchmarks designed to evaluate {\color{black}embodied} IBSA's performance.
Finally, {\color{black}challenges, future directions, and concluding remarks} are presented in Section~\ref{sec_open} and Section~\ref{sec_conclusion}, respectively.

\section{Evolution of Base Station: From Single-Functional Communication Infrastructure to Multi-Functional LAM-Enabled IBSA}\label{sec_evolution}
The transition of BS from traditional communication infrastructure to the LAM-enabled IBSA represents a fundamental paradigm shift. 
This transformation is not merely functional but is mirrored by a profound physical evolution of the BS hardware itself. 
Over decades, BSs have evolved from centralized analog cabinets with passive antennas to distributed, integrated, and multi-modal intelligent agents. 
Fig.~\ref{fig_BS} illustrates this physical evolutionary path, which provides the hardware foundation for the LAM-enabled IBSA that integrates perception, communication, and computation in the 6G era and is central to this paper.
\subsection{Large Model in Communication Network}
LAMs are pivotal in driving the evolution of 6G networks due to their versatility and generalization abilities across diverse tasks. Their capacity for cross-task generalization means that these models can be applied to multiple network scenarios without the need for extensive retraining, making them invaluable in dynamic environments~\cite{ref_LAM_2}. Furthermore, their proficiency in few-shot learning allows for efficient adaptation to new tasks with minimal data, reducing the need for large-scale training datasets~\cite{ref_LAM_3}. Additionally, {\color{black}LAMs} exhibit emergent abilities, where they can develop novel problem-solving strategies that were not explicitly part of their training data~\cite{ref_LAM_1}.  

The intersection of large models and edge learning-based agentic models holds immense potential for revolutionizing communication networks. By deploying these models on edge devices, telecommunications systems can enhance both the intelligence and efficiency of resource allocation and management. Research on AI-native telecommunications networks reveals how large models, particularly foundation models, can empower edge devices to handle complex tasks autonomously~\cite{ref_LAM_2}. This enables real-time decision-making, reducing latency and improving network reliability. Notably, agentic AI models integrated within telecommunications systems allow BSs to act as intelligent agents that learn from interactions with the environment, leading to self-optimizing networks capable of anticipating and responding to various network conditions~\cite{ref_LAM_3}.
\begin{table*}[!t]	
	\centering	
	\caption{Representative multi-modality dataset in communication domain. $\triangle$ denotes uncertainty.}
	\label{table_dataset}
	\begin{small}
		\begin{tabular}{|c|c:c:c:c:c:c|c:c:c:c:c|c:c:c:c|c|c|c|}
			\hline
			\multirow{2}{*}{\textbf{Ref.}\vspace{-2mm}} & \multicolumn{6}{c|}{\textbf{Platform}} & \multicolumn{5}{c|}{\textbf{Sensors/Modality}} & \multicolumn{4}{c|}{\textbf{Weather/Time}} & \multicolumn{1}{c|}{\multirow{2}{*}{\begin{tabular}[c]{@{}c@{}} \textbf{Multi-}\\ \textbf{Scenario}\vspace{-2mm}\end{tabular}}} & \multicolumn{1}{c|}{\multirow{2}{*}{\begin{tabular}[c]{@{}c@{}}\textbf{Customized}\\ \textbf{Definition}\vspace{-2mm}\end{tabular}}} & \multicolumn{1}{c|}{\multirow{2}{*}{\begin{tabular}[c]{@{}c@{}}\textbf{Aerial}\\ \textbf{Target}\vspace{-2mm}\end{tabular}}} \\ \cline{2-16}
			& \multicolumn{1}{c:}{CARLA} & \multicolumn{1}{c:}{Sionna} & \multicolumn{1}{c:}{\begin{tabular}[c]{@{}c@{}}Wireless\\ InSite\end{tabular}} & \multicolumn{1}{c:}{\begin{tabular}[c]{@{}c@{}}Unreal\\ Engine\end{tabular}} & \multicolumn{1}{c:}{\begin{tabular}[c]{@{}c@{}}Real\\ World\end{tabular}} & \multicolumn{1}{c|}{Others} & \multicolumn{1}{l:}{Camera} & \multicolumn{1}{l:}{lidar} & \multicolumn{1}{l:}{Radar} & \multicolumn{1}{l:}{CSI} & \multicolumn{1}{l|}{IR} & \multicolumn{1}{l:}{Rain} & \multicolumn{1}{l:}{Snow} & \multicolumn{1}{c:}{Fog} & \multicolumn{1}{l|}{Night} & \multicolumn{1}{c|}{} & \multicolumn{1}{c|}{} & \multicolumn{1}{c|}{} \\ \hline
			\cite{ref_DeepSense}& {\color{red}\ding{54}} & {\color{red}\ding{54}} & {\color{red}\ding{54}} & {\color{red}\ding{54}} & {\color{green}\ding{51}} & {\color{red}\ding{54}} & {\color{green}\ding{51}} & {\color{green}\ding{51}} & {\color{green}\ding{51}} & {\color{green}\ding{51}} & {\color{red}\ding{54}} & {\color{green}\ding{51}} & {\color{red}\ding{54}} & {\color{red}\ding{54}} & {\color{green}\ding{51}} & {\color{green}\ding{51}} & {\color{red}\ding{54}} & {\color{green}\ding{51}} \\ \hdashline
			\cite{ref_DeepVerse} & {\color{green}\ding{51}} & {\color{red}\ding{54}} & {\color{green}\ding{51}} & {\color{red}\ding{54}} & {\color{red}\ding{54}} & {\color{red}\ding{54}} & {\color{green}\ding{51}} & {\color{green}\ding{51}} & {\color{green}\ding{51}} & {\color{green}\ding{51}} & {\color{red}\ding{54}} & {\color{red}\ding{54}} & {\color{red}\ding{54}} & {\color{red}\ding{54}} & {\color{red}\ding{54}} & {\color{green}\ding{51}} & {\color{red}\ding{54}} & {\color{red}\ding{54}} \\ \hdashline
			\cite{ref_Unreal} & {\color{red}\ding{54}} & {\color{red}\ding{54}} & {\color{red}\ding{54}} & {\color{green}\ding{51}} & {\color{red}\ding{54}} & {\color{red}\ding{54}} & {\color{green}\ding{51}} & {\color{green}\ding{51}} & {\color{green}\ding{51}} & {\color{green}\ding{51}} & {\color{red}\ding{54}} & $ \triangle $ & $ \triangle $ & $ \triangle $ & $ \triangle $ & {\color{green}\ding{51}} & {\color{green}\ding{51}} & {\color{green}\ding{51}} \\ \hdashline
			\cite{ref_MVX} & {\color{green}\ding{51}} & {\color{green}\ding{51}} & {\color{red}\ding{54}} & {\color{red}\ding{54}} & {\color{red}\ding{54}} & {\color{red}\ding{54}} & {\color{green}\ding{51}} & {\color{green}\ding{51}} & {\color{red}\ding{54}} & {\color{green}\ding{51}} & {\color{red}\ding{54}} & $ \triangle $ & $ \triangle $ & $ \triangle $ & $ \triangle $ & {\color{green}\ding{51}} & {\color{green}\ding{51}} & {\color{red}\ding{54}} \\ \hdashline
			\cite{ref_M3SC} & {\color{red}\ding{54}} & {\color{red}\ding{54}} & {\color{green}\ding{51}} & {\color{red}\ding{54}} & {\color{red}\ding{54}} & \begin{tabular}[c]{@{}l@{}} \text{\ \ \,}AirSim,\\ WaveFarer\end{tabular} & {\color{green}\ding{51}} & {\color{green}\ding{51}} & {\color{green}\ding{51}} & {\color{green}\ding{51}} & {\color{red}\ding{54}} & {\color{green}\ding{51}} & {\color{green}\ding{51}} & {\color{green}\ding{51}} & {\color{green}\ding{51}} & {\color{red}\ding{54}} & {\color{green}\ding{51}} & {\color{red}\ding{54}} \\ \hdashline
			\cite{ref_SoM} & {\color{red}\ding{54}} & {\color{red}\ding{54}} & {\color{green}\ding{51}} & {\color{red}\ding{54}} & {\color{red}\ding{54}} & \begin{tabular}[c]{@{}l@{}}\text{\ \ \,}AirSim,\\ WaveFarer\end{tabular} & {\color{green}\ding{51}} & {\color{green}\ding{51}} & {\color{green}\ding{51}} & {\color{green}\ding{51}} & {\color{red}\ding{54}} & {\color{green}\ding{51}} & {\color{green}\ding{51}} & {\color{red}\ding{54}} & {\color{green}\ding{51}} & {\color{red}\ding{54}} & {\color{green}\ding{51}} & {\color{red}\ding{54}} \\ \hdashline
			\cite{ref_TVT_dataset} & {\color{red}\ding{54}} & {\color{red}\ding{54}} & {\color{green}\ding{51}} & {\color{red}\ding{54}} & {\color{green}\ding{51}} & \begin{tabular}[c]{@{}l@{}}Blender,\\ Blensor\end{tabular} & {\color{green}\ding{51}} & {\color{green}\ding{51}} & {\color{red}\ding{54}} & {\color{red}\ding{54}} & {\color{red}\ding{54}} & {\color{red}\ding{54}} & {\color{red}\ding{54}} & {\color{red}\ding{54}} & {\color{red}\ding{54}} & {\color{green}\ding{51}} & {\color{red}\ding{54}} & {\color{red}\ding{54}} \\ \hdashline
			\cite{ref_Infocom_Dataset} & {\color{red}\ding{54}} & {\color{red}\ding{54}} & {\color{red}\ding{54}} & {\color{red}\ding{54}} & {\color{green}\ding{51}} & {\color{red}\ding{54}} & {\color{red}\ding{54}} & {\color{green}\ding{51}} & {\color{red}\ding{54}} & {\color{red}\ding{54}} & {\color{red}\ding{54}} & {\color{red}\ding{54}} & {\color{red}\ding{54}} & {\color{red}\ding{54}} & {\color{red}\ding{54}} & {\color{green}\ding{51}} & {\color{red}\ding{54}} & {\color{red}\ding{54}} \\ \hdashline
			\cite{ref_TMC_Dataset} & {\color{red}\ding{54}} & {\color{red}\ding{54}} & {\color{green}\ding{51}} & {\color{red}\ding{54}} & {\color{green}\ding{51}} & \begin{tabular}[c]{@{}l@{}}Blender,\\ Blensor\end{tabular} & {\color{green}\ding{51}} & {\color{green}\ding{51}} & {\color{red}\ding{54}} & {\color{red}\ding{54}} & {\color{red}\ding{54}} & {\color{red}\ding{54}} & {\color{red}\ding{54}} & {\color{red}\ding{54}} & {\color{red}\ding{54}} & {\color{green}\ding{51}} & {\color{green}\ding{51}} & {\color{red}\ding{54}} \\ \hdashline
			\cite{ref_WC_Dataset} & {\color{red}\ding{54}} & {\color{red}\ding{54}} & {\color{red}\ding{54}} & {\color{red}\ding{54}} & {\color{green}\ding{51}} & {\color{red}\ding{54}} & {\color{green}\ding{51}} & {\color{green}\ding{51}} & {\color{green}\ding{51}} & {\color{green}\ding{51}} &{\color{green}\ding{51}}& {\color{red}\ding{54}} & {\color{red}\ding{54}} & {\color{red}\ding{54}} & {\color{red}\ding{54}} & {\color{green}\ding{51}} & {\color{red}\ding{54}} & {\color{red}\ding{54}} \\ \hdashline
			\cite{ref_HADAR} & {\color{red}\ding{54}} & {\color{red}\ding{54}} & {\color{red}\ding{54}} & {\color{red}\ding{54}} & {\color{green}\ding{51}} & \begin{tabular}[c]{@{}l@{}}Blender,\\ OpenGL\end{tabular} & {\color{green}\ding{51}} & {\color{green}\ding{51}} & {\color{red}\ding{54}} & {\color{red}\ding{54}} &{\color{green}\ding{51}}& {\color{red}\ding{54}} & {\color{red}\ding{54}} & {\color{red}\ding{54}} & {\color{red}\ding{54}} & {\color{green}\ding{51}} & {\color{green}\ding{51}} & {\color{red}\ding{54}} \\ \hdashline
			\cite{yang2024v2x} & {\color{red}\ding{54}} & {\color{red}\ding{54}} & {\color{red}\ding{54}} & {\color{red}\ding{54}} & {\color{green}\ding{51}} & {\color{red}\ding{54}} & {\color{green}\ding{51}} & {\color{green}\ding{51}} & {\color{green}\ding{51}} & {\color{red}\ding{54}} &{\color{green}\ding{51}} &{\color{green}\ding{51}} & {\color{green}\ding{51}} &{\color{green}\ding{51}} & {\color{green}\ding{51}} & {\color{red}\ding{54}} & {\color{red}\ding{54}} & {\color{red}\ding{54}} \\ \hdashline
			\cite{mao2025multimodal} & {\color{green}\ding{51}} & {\color{green}\ding{51}} & {\color{red}\ding{54}} & {\color{red}\ding{54}} & {\color{red}\ding{54}} & Blender & {\color{green}\ding{51}} & {\color{green}\ding{51}} & {\color{green}\ding{51}} & {\color{green}\ding{51}} &{\color{red}\ding{54}} &{\color{green}\ding{51}} & {\color{green}\ding{51}} &{\color{green}\ding{51}} & {\color{red}\ding{54}} & {\color{green}\ding{51}} & {\color{red}\ding{54}} & {\color{red}\ding{54}} \\ \hline
		\end{tabular}
	\end{small}
\end{table*}

	\begin{table*}[!t]		
		\centering
		\caption{Representative studies of multimodal perception fusion for communication systems.}
		\label{table_app}
		\begin{small}
			\begin{tabular}{|c|c:c:c:c:c:c|c:c:c:c:c|c:c:c:c|}
				\hline
				\multirow{2}{*}{\textbf{Ref.}\vspace{-8mm}} & \multicolumn{6}{c|}{\textbf{Platform}} & \multicolumn{5}{c|}{\textbf{Sensors/Modality}} & \multicolumn{4}{c|}{\textbf{Application}} \\ \cline{2-16}	
				& \multicolumn{1}{c:}{CARLA} & \multicolumn{1}{c:}{Sionna} & \multicolumn{1}{c:}{\begin{tabular}[c]{@{}c@{}}Wireless\\ InSite\end{tabular}} & \multicolumn{1}{c:}{\begin{tabular}[c]{@{}c@{}}Unreal\\ Engine\end{tabular}} & \multicolumn{1}{c:}{\begin{tabular}[c]{@{}c@{}}Real\\ World\end{tabular}} & \multicolumn{1}{c|}{Others} & \multicolumn{1}{l:}{Camera} & \multicolumn{1}{l:}{lidar} & \multicolumn{1}{l:}{Radar} & \multicolumn{1}{l:}{CSI} & \multicolumn{1}{l|}{IR} & \multicolumn{1}{l:}{\begin{tabular}[c]{@{}c@{}}Detection,\\ Localization\end{tabular}} & \multicolumn{1}{c:}{\begin{tabular}[c]{@{}c@{}}Beam\\ Prediction\end{tabular}} & \multicolumn{1}{c:}{\begin{tabular}[c]{@{}c@{}}Antenna\\ Position\\ Optimization\end{tabular}} & \multicolumn{1}{c|}{\begin{tabular}[c]{@{}c@{}}Channel\\ Prediction\end{tabular}} \\ \hline
				\cite{ref_Unreal}& {\color{red}\ding{54}} & {\color{red}\ding{54}} & {\color{red}\ding{54}} & {\color{green}\ding{51}} & {\color{red}\ding{54}} & {\color{red}\ding{54}} & {\color{green}\ding{51}} & {\color{green}\ding{51}} & {\color{green}\ding{51}} & {\color{green}\ding{51}} & {\color{red}\ding{54}} & {\color{green}\ding{51}} & {\color{red}\ding{54}} & {\color{red}\ding{54}} & {\color{red}\ding{54}} \\ \hdashline
				\cite{ref_MVX}& {\color{green}\ding{51}} & {\color{green}\ding{51}} & {\color{red}\ding{54}} & {\color{red}\ding{54}} & {\color{red}\ding{54}} & {\color{red}\ding{54}} & {\color{green}\ding{51}} & {\color{green}\ding{51}} & {\color{red}\ding{54}} & {\color{green}\ding{51}} & {\color{red}\ding{54}} & {\color{green}\ding{51}} & {\color{green}\ding{51}} & {\color{green}\ding{51}} & {\color{red}\ding{54}} \\ \hdashline
				\cite{ref_SoM}& {\color{red}\ding{54}} & {\color{red}\ding{54}} & {\color{green}\ding{51}} & {\color{red}\ding{54}} & {\color{red}\ding{54}} & \begin{tabular}[c]{@{}l@{}}{\ \ \,}AirSim,\\ WaveFarer\end{tabular} & {\color{green}\ding{51}} & {\color{green}\ding{51}} & {\color{green}\ding{51}} & {\color{green}\ding{51}} & {\color{red}\ding{54}} & {\color{red}\ding{54}} & {\color{red}\ding{54}} & {\color{red}\ding{54}} & {\color{green}\ding{51}} \\ \hdashline
				\cite{ref_TVT_dataset}& {\color{red}\ding{54}} & {\color{red}\ding{54}} & {\color{green}\ding{51}} & {\color{red}\ding{54}} & {\color{green}\ding{51}} & \begin{tabular}[c]{@{}l@{}}Blender,\\ Blensor\end{tabular} & {\color{green}\ding{51}} & {\color{green}\ding{51}} & {\color{red}\ding{54}} & {\color{red}\ding{54}} & {\color{red}\ding{54}} & {\color{red}\ding{54}} & {\color{green}\ding{51}} & {\color{red}\ding{54}} & {\color{red}\ding{54}} \\ \hdashline
				\cite{ref_Infocom_Dataset}& {\color{red}\ding{54}} & {\color{red}\ding{54}} & {\color{red}\ding{54}} & {\color{red}\ding{54}} & {\color{green}\ding{51}} & {\color{red}\ding{54}} & {\color{red}\ding{54}} & {\color{green}\ding{51}} & {\color{red}\ding{54}} & {\color{red}\ding{54}} & {\color{red}\ding{54}} & {\color{red}\ding{54}} & {\color{green}\ding{51}} & {\color{red}\ding{54}} & {\color{red}\ding{54}} \\ \hdashline
				\cite{ref_TMC_Dataset}& {\color{red}\ding{54}} & {\color{red}\ding{54}} & {\color{green}\ding{51}} & {\color{red}\ding{54}} & {\color{green}\ding{51}} & \begin{tabular}[c]{@{}l@{}}Blender,\\ Blensor\end{tabular} & {\color{green}\ding{51}} & {\color{green}\ding{51}} & {\color{red}\ding{54}} & {\color{red}\ding{54}} & {\color{red}\ding{54}} & {\color{red}\ding{54}} & {\color{green}\ding{51}} & {\color{red}\ding{54}} & {\color{red}\ding{54}} \\ \hdashline
				\cite{ref_WC_Dataset}& {\color{red}\ding{54}} & {\color{red}\ding{54}} & {\color{red}\ding{54}} & {\color{red}\ding{54}} & {\color{green}\ding{51}} & {\color{red}\ding{54}} & {\color{green}\ding{51}} & {\color{green}\ding{51}} & {\color{green}\ding{51}} & {\color{green}\ding{51}} & {\color{green}\ding{51}} & {\color{green}\ding{51}} & {\color{green}\ding{51}} & {\color{red}\ding{54}} & {\color{red}\ding{54}} \\ \hdashline
				\cite{ref_HADAR}& {\color{red}\ding{54}} & {\color{red}\ding{54}} & {\color{red}\ding{54}} & {\color{red}\ding{54}} & {\color{green}\ding{51}} & \begin{tabular}[c]{@{}l@{}}Blender,\\ OpenGL\end{tabular} & {\color{green}\ding{51}} & {\color{green}\ding{51}} & {\color{red}\ding{54}} & {\color{red}\ding{54}} & {\color{green}\ding{51}} & {\color{green}\ding{51}} & {\color{red}\ding{54}} & {\color{red}\ding{54}} & {\color{red}\ding{54}} \\ \hdashline
				\cite{ref_24_arXiv_DeepSense_Predict} & {\color{red}\ding{54}} & {\color{red}\ding{54}} & {\color{red}\ding{54}} & {\color{red}\ding{54}} & {\color{green}\ding{51}} & {\color{red}\ding{54}} & {\color{green}\ding{51}} & {\color{red}\ding{54}} & {\color{red}\ding{54}} & {\color{red}\ding{54}} & {\color{red}\ding{54}} & {\color{red}\ding{54}} & {\color{green}\ding{51}} & {\color{red}\ding{54}} & {\color{red}\ding{54}} \\ \hdashline
				\cite{ref_25_arXiv_carla_sionna} & {\color{green}\ding{51}} & {\color{red}\ding{54}} & {\color{red}\ding{54}} & {\color{red}\ding{54}} & {\color{red}\ding{54}} & MATLAB & {\color{green}\ding{51}} & {\color{green}\ding{51}} & {\color{green}\ding{51}} & {\color{red}\ding{54}} & {\color{red}\ding{54}} & {\color{red}\ding{54}} & {\color{green}\ding{51}} & {\color{red}\ding{54}} & {\color{red}\ding{54}} \\ \hdashline
				\cite{ref_VTCFall_carla} & {\color{green}\ding{51}} & {\color{red}\ding{54}} & {\color{green}\ding{51}} & {\color{red}\ding{54}} & {\color{red}\ding{54}} & {\color{red}\ding{54}} & {\color{green}\ding{51}} & {\color{green}\ding{51}} & {\color{green}\ding{51}} & {\color{green}\ding{51}} & {\color{red}\ding{54}} & {\color{red}\ding{54}} & {\color{green}\ding{51}} & {\color{red}\ding{54}} & {\color{red}\ding{54}} \\ \hdashline
				\cite{ref_TMLCN} & {\color{green}\ding{51}} & {\color{green}\ding{51}} & {\color{red}\ding{54}} & {\color{red}\ding{54}} & {\color{red}\ding{54}} & {\color{red}\ding{54}} & {\color{green}\ding{51}} & {\color{green}\ding{51}} & {\color{red}\ding{54}} & {\color{green}\ding{51}} & {\color{red}\ding{54}} & {\color{red}\ding{54}} & {\color{red}\ding{54}} & {\color{red}\ding{54}} & {\color{green}\ding{51}} \\ \hdashline
				\cite{ref_TN}& {\color{red}\ding{54}} & {\color{red}\ding{54}} & {\color{green}\ding{51}} & {\color{red}\ding{54}} & {\color{red}\ding{54}} & {\color{red}\ding{54}} & {\color{green}\ding{51}} & {\color{green}\ding{51}} & {\color{red}\ding{54}} & {\color{red}\ding{54}} & {\color{red}\ding{54}} & {\color{red}\ding{54}} & {\color{green}\ding{51}} & {\color{red}\ding{54}} & {\color{red}\ding{54}} \\ \hline
			\end{tabular}
		\end{small}
	\end{table*}
	
	\subsection{Large Model Enabled Embodied Intelligent Base Station Agent}
	Embodied intelligence refers to systems where intelligence is not just about processing information but is tightly coupled with physical action, perception, and learning within an environment. Traditionally, embodied intelligence has been applied to robotics, where it involves the dynamic interaction of sensors, actuators, and perception to enable real-time adaptation to physical environments~\cite{ref_Embodied_1, ref_Embodied_2}. The interaction between the agent and its surroundings forms a continuous loop of sensory input, decision-making, and motor output, providing agents with the ability to learn and adapt~\cite{ref_Embodied_3}.
	
	However, this paradigm is now extending beyond robotics and into the realm of communication infrastructure. As networks become increasingly intelligent, embodied intelligence concepts are being integrated into network systems, particularly in 6G environments. This shift allows BSs to act as ``embodied agents" that not only process data but actively sense and adjust to the physical network environment, improving network performance through adaptive actions like {\color{black}beamforming, resource allocation, and optimization of spectrum}~\cite{ref_Embodied_2}. This transformation is akin to how autonomous vehicles use multi-modal sensor fusion to navigate their surroundings~\cite{ref_Embodied_1}. In a similar manner, network infrastructure can leverage multi-modal sensory data (e.g., radio-frequency perception, visual data) to improve decision-making in dynamic wireless environments~\cite{ref_Embodied_3}.
	Ultimately, the convergence of embodied intelligence and communication networks has the potential to create more adaptive, resilient systems that are capable of real-time optimization and task-specific decision-making. Such integration marks a fundamental shift in how networks operate, moving from passive infrastructure to intelligent {\color{black}and} autonomous systems that can physically interact with their environment~\cite{ref_Embodied_2}.
	
	Treating BSs as {\color{black}embodied} intelligent BS agents (IBSAs) is an innovative approach in 6G networks. This perspective transforms BSs from mere communication relays to perceptive entities capable of actively perception and adapting to their physical environments. With the integration of sensors and AI, BSs can optimize network performance, such as adjusting beamforming based on real-time conditions. However, this shift presents challenges, including the need for seamless multi-modal data fusion and real-time decision-making capabilities that balance computational resources with the network's dynamic demands.
	Building upon the vision of BSs as {\color{black}embodied} IBSAs, there exists two compelling application scenarios: autonomous driving and low-altitude safety. In cooperative vehicle-infrastructure autonomous driving, {\color{black}embodied} IBSAs with integrated perception capabilities can play a pivotal role in supporting vehicle-to-everything (V2X) communications through multi-modal data fusion. This allows for real-time, context-aware decision-making that enhances vehicle navigation in complex environments. Similarly, in low-altitude airspace, {\color{black}embodied} IBSAs can actively detect and neutralize potential uncrewed aerial vehicle (UAV) threats. As we move towards 6G, these intelligent, adaptive systems will be critical for ensuring both the safety and efficiency of future networks. 
	
	\subsection{Embodied Intelligent Base Station Agent Enabled Autonomous Driving and Low-Altitude Safety}
	Data is of paramount importance for AI research. However, obtaining real-world data and annotations for cooperative autonomous driving and low-altitude safety scenarios is often exceedingly challenging~\cite{yang2024v2x}.
	Thanks to advances in autonomous driving research, a number of mature simulation environments now support algorithm development and validation within digital twins, including Unreal Engine (a cross‑platform, highly extensible real-time 3D engine)~\cite{ref_UE}, CARLA (an open‑source autonomous driving simulator)~\cite{ref_carla}, Blender (a free, open‑source, cross‑platform 3D creation suite)~\cite{ref_blender}, Blensor (a free, open‑source simulation package for lidar and Kinect sensors)~\cite{ref_blensor}, AirSim (a simulator for drones, cars, and more, built on Unreal Engine)~\cite{ref_AirSim}, and OpenGL (an open, cross‑platform, cross‑language graphics rendering application programming interface (API))~\cite{ref_OpenGL}.
	
	Similarly, the wireless‑signal‑processing community benefits from its own set of digital‑twin tools that lay the groundwork for {\color{black}embodied} IBSA: Wireless Insite (a predictive tool for understanding wireless coverage, channel multipath, and data throughput)~\cite{ref_WirelessInSite}, Sionna (a hardware‑accelerated, differentiable open‑source library for communications research)~\cite{ref_sionna}, and WaveFarer (a high‑fidelity radar simulator)~\cite{ref_WaveFarer}.
	
	Meanwhile, autonomous driving has produced numerous publicly available multimodal datasets~\cite{ref_24_Bosch_radar,ref_24_TIV_radar_camera}.
	Recent surveys, such as~\cite{ref_24_TITS_lidar_camera} on lidar and camera fusion, and~\cite{ref_24_TIV_radar_camera} on radar and camera fusion, provide comprehensive overviews, while works like~\cite{ref_23_ICRA_lidar_camera},~\cite{ref_25_TITS_radar_camera}, and~\cite{ref_24_IROS_radar_camera} push the state of the art in sensor fusion.
	
	Methods for multimodal fusion in autonomous driving offer valuable insights for wireless signal processing.  Traditional communications rely on wireless pilot signals alone to estimate the channel and perform demodulation~\cite{ref_gz_1,ref_gz_2,ref_gz_3}, but ISAC has emerged as a core 6G capability in which enhanced perception and communications can mutually reinforce one another, enabling applications such as cellular positioning, low‑altitude safety, and autonomous driving~\cite{ref_LZR_JSAC,ref_lzr_JSTSP,ref_lzr_IoTJ,ref_lzr_Network,ref_zxy_JSAC}.
	Table~\ref{table_dataset} presents an overview of widely used multimodal datasets that serve as the foundation for perception research in the communications domain.
	Building upon these datasets, Table~\ref{table_app} summarizes representative multimodal perception approaches, including those based on proprietary simulations, real-world measurements, and high-quality public data.
	
	In light of current trends toward large‑scale models and embodied intelligence, and building on the trajectory of future communications, the capabilities of existing simulators and datasets, and the wealth of multimodal perception schemes in the literature, this paper explores how {\color{black}embodied} IBSA can be most effectively applied in two key areas: vehicle-road cooperative autonomous driving and ubiquitous base‑station-enabled low‑altitude safety.
	
		\subsection{Comparison with Existing Next-Generation Base Station Architectures}
		
		Recent advances at the intersection of LAMs and 6G have led to several architectural proposals that are closely related to, but still fundamentally different from, the IBSA architecture considered in this paper. Broadly speaking, existing work can be grouped into three lines: (i) LLM/LAM-centric frameworks that endow telecom systems with powerful cognitive capabilities; (ii) network- and edge-centric architectures that distribute large models across a 6G computing continuum; and (iii) agentic or embodied network architectures that explicitly consider interaction with the physical world. This subsection clarifies how IBSA relates to and differs from these lines of research.
		
		The first line focuses on large models as telecom-specific ``brains", but does not explicitly architect a next-generation BS. TelecomGPT is a representative example: it builds a domain-specialized LLM by constructing a telecom corpus, designing pre-training and instruction-tuning pipelines, and defining benchmark tasks such as standard interpretation, troubleshooting, and radio access network configuration assistance~\cite{ref_LAM_2}. Similarly, the comprehensive survey of LAMs for future communications overviews how LAMs can support a broad spectrum of communication tasks, from channel prediction and beam management to resource allocation and network automation, and sketches a high-level reference framework where LAMs provide AI services to the network~\cite{jiang2025comprehensive}. These works clearly articulate the role of LLMs/LAMs as cognitive engines, but the BS is mainly treated as a logical endpoint or data source that queries the model. They do not describe how a single BS should internally organise multimodal perception, LAM-based cognition, and executable actions, nor do they consider safety-critical, closed-loop control at the BS level.
		
		The second line emphasises the network and edge-computing perspective of deploying LAMs. The wireless large AI model framework discusses how a unified LAM can interface with multiple radio functions, proposes an AI-native 6G network architecture, and identifies enabling technologies such as model compression, parallelism, and knowledge distillation~\cite{ref_LAM_3}. Edge LAM architectures go further by decomposing LAMs and distributing their submodules across geographically separated edge devices, thereby addressing communication, computation, and storage bottlenecks while enabling general-purpose intelligent services~\cite{ref_LAM_1}. The 6G white paper on LLMs in the computing continuum complements these works by positioning LLMs within a cloud-edge-device hierarchy, where BSs act as part of the edge layer that hosts or accesses LLM-based services and participates in cross-layer orchestration~\cite{abel2025large}. Taken together, these works provide valuable guidance on where large models should reside, how they can be split, and how they interact with edge resources. However, they largely view the BS as an element of the computing fabric; the internal BS architecture is not the focus, and explicit integration of perception, communication, and local actuation into an embodied BS agent is outside their scope.
		
		The third line explicitly moves from ``large models" to ``agents" and from passive observation to embodiment. F. Jiang \textit{et al.} in~\cite{jiang2025large} systematically introduced agentic stacks built on top of LAMs, including planning, tool use, memory, and feedback mechanisms, and envisages multi-agent systems for network operation, optimization, and service provisioning. The recently proposed wireless embodied large AI further argues that large models for wireless systems should close the loop with the physical environment via active environmental probing, on-line learning, and interaction with physical interfaces; it outlines design principles and a generic system structure for embodied wireless entities such as user devices or BSs~\cite{ref_Embodied_2}. In parallel, the survey on ISAC towards multifunctional perceptive networks shows how BSs may evolve into perceptive nodes that jointly perform communication and perception by sharing spectrum, hardware, and signal-processing pipelines, and proposes several perceptive-network architectures~\cite{cui2025integrated}. These works are highly complementary to IBSA: they highlight the importance of embodiment, agentic behaviour, and integrated perception. Nevertheless, they are either formulated at the level of generic wireless entities or at the network/cluster level, without providing a concrete, BS-centric hardware-software architecture that unifies multimodal perception, LAM-based cognition, and BS-side actuation in specific safety-critical scenarios.

		\begin{table*}[!t]
			\centering
			\caption{Systematic comparison between representative next-generation BS-related large-model/ISAC architectures and the proposed large-model-enabled IBSA.}
			\label{table_cmp}
			\begin{tabular}{|c|c:c:c|c:c|c:c|c:c|} 
				\hline
				\multirow{2}{*}{\textbf{Ref.}} & \multicolumn{3}{c|}{\begin{tabular}[c]{@{}c@{}}\textbf{Functional Focus of}\\ \textbf{Next-generation BSs}\end{tabular}} & \multicolumn{2}{c|}{\begin{tabular}[c]{@{}c@{}}\textbf{Core Intelligence}\\ \textbf{at the BS}\end{tabular}} & \multicolumn{2}{c|}{\begin{tabular}[c]{@{}c@{}}\textbf{BS-environment Interaction}\\ \textbf{via Wireless Signals}\end{tabular}} & \multicolumn{2}{c|}{\begin{tabular}[c]{@{}c@{}}\textbf{Key Scenarios for} \\ \textbf{Next-generation BSs}\end{tabular}} \\ \cline{2-10} 
				& \multicolumn{1}{c:}{{Comm.}} & \multicolumn{1}{c:}{{Perception}} & \multicolumn{1}{c|}{{Computation}} & \multicolumn{1}{c:}{\text{\ \ {LAM}\ \ } } & \multicolumn{1}{c|}{{LLM}} & \multicolumn{1}{c:}{\begin{tabular}[c]{@{}c@{}}Passive Reception \\ Only\end{tabular}} & \multicolumn{1}{c|}{{Active Interaction}} & \multicolumn{1}{c:}{\begin{tabular}[c]{@{}c@{}}{Autonomous}\\{ Driving}\end{tabular}} & \multicolumn{1}{c|}{\begin{tabular}[c]{@{}c@{}}{Low-altitude}\\ {Safety}\end{tabular}} \\ \hline
				\cite{ref_LAM_2}& {\color{green}\ding{51}} & {\color{red}\ding{54}} & {\color{red}\ding{54}} & {\color{red}\ding{54}} & {\color{green}\ding{51}} & {\color{green}\ding{51}} & {\color{red}\ding{54}} & {\color{red}\ding{54}} & {\color{red}\ding{54}} \\ \hdashline
				\cite{ref_LAM_3}& {\color{green}\ding{51}} & {\color{green}\ding{51}} & {\color{green}\ding{51}} & {\color{green}\ding{51}} & {\color{red}\ding{54}} & {\color{green}\ding{51}} & {\color{red}\ding{54}} & {\color{red}\ding{54}} & {\color{red}\ding{54}} \\ \hdashline
				\cite{ref_LAM_1}& {\color{green}\ding{51}} & {\color{green}\ding{51}} & {\color{green}\ding{51}} & {\color{green}\ding{51}} & {\color{red}\ding{54}} & {\color{green}\ding{51}} & {\color{red}\ding{54}} & {\color{red}\ding{54}} & {\color{red}\ding{54}} \\ \hdashline
				\cite{ref_Embodied_2}& {\color{green}\ding{51}} & {\color{green}\ding{51}} & {\color{green}\ding{51}} & {\color{green}\ding{51}} & {\color{red}\ding{54}} & {\color{red}\ding{54}} & {\color{green}\ding{51}} & {\color{green}\ding{51}} & {\color{red}\ding{54}} \\ \hdashline
				\cite{jiang2025comprehensive}& {\color{green}\ding{51}} & {\color{red}\ding{54}} & {\color{green}\ding{51}} & {\color{green}\ding{51}} & {\color{red}\ding{54}} & {\color{green}\ding{51}} & {\color{red}\ding{54}} & {\color{red}\ding{54}} & {\color{red}\ding{54}} \\ \hdashline
				\cite{abel2025large}& {\color{green}\ding{51}} & {\color{red}\ding{54}} & {\color{green}\ding{51}} & {\color{red}\ding{54}} & {\color{green}\ding{51}} & {\color{green}\ding{51}} & {\color{red}\ding{54}} & {\color{green}\ding{51}} & {\color{red}\ding{54}} \\ \hdashline
				\cite{jiang2025large}& {\color{green}\ding{51}} & {\color{red}\ding{54}} & {\color{green}\ding{51}} & {\color{green}\ding{51}} & {\color{red}\ding{54}} & {\color{green}\ding{51}} & {\color{red}\ding{54}} & {\color{red}\ding{54}} & {\color{red}\ding{54}} \\ \hdashline
				\cite{cui2025integrated}& {\color{green}\ding{51}} & {\color{green}\ding{51}} & {\color{green}\ding{51}} & {\color{red}\ding{54}} & {\color{red}\ding{54}} & {\color{red}\ding{54}} & {\color{green}\ding{51}} & {\color{green}\ding{51}} & {\color{green}\ding{51}} \\ \hdashline
				\textbf{Ours}& {\color{green}\ding{51}} & {\color{green}\ding{51}} & {\color{green}\ding{51}} & {\color{green}\ding{51}} & {\color{red}\ding{54}} & {\color{red}\ding{54}} & {\color{green}\ding{51}} & {\color{green}\ding{51}} & {\color{green}\ding{51}} \\ \hline
			\end{tabular}
		\end{table*}
		
		In contrast to the above, the proposed large-model-enabled IBSA architecture is explicitly BS-centric and scenario-driven. Each BS is treated as an embodied agent whose functional focus simultaneously covers communication, perception, and computation. An LAM serves as the unified cognitive core at the BS, while multimodal RF, visual, and point-cloud perception modules feed into this core, and the BS exerts actions through both traditional control-plane commands (e.g., beamforming, power control, scheduling) and environment-shaping wireless interactions (e.g., cooperative perception with vehicles and uncrewed aerial vehicle (UAVs), spectrum-level intervention for low-altitude safety). The architecture is instantiated through a three-layer perception-cognition-execution pipeline, coupled with cloud-edge-end collaboration, digital twins for data generation and simulation, and communication-computation co-design. Unlike purely LAM-centric surveys, IBSA therefore provides a concrete blueprint for how a single BS can be upgraded into an embodied, large-model-enabled agent in specific safety-critical use cases such as cooperative vehicle-road autonomous driving and low-altitude security.
		
		At the same time, IBSA does not aim to replace the above frameworks; rather, it builds on and complements them. Compared with TelecomGPT and LAM surveys~\cite{ref_LAM_2,jiang2025comprehensive}, IBSA adds explicit multimodal perception and embodiment at the BS level. Compared with edge/continuum LAM architectures~\cite{ref_LAM_1,ref_LAM_3,abel2025large}, IBSA assumes that many of their techniques, e.g., model decomposition, offloading, and lifecycle management, will be applied under the hood to support the large-model brain of each BS, but these system-level issues are not the main focus of this paper. Compared with WELAI~\cite{ref_Embodied_2}, agentic-AI~\cite{jiang2025large}, and ISAC-based perceptive-network~\cite{cui2025integrated} frameworks, IBSA specialises their general principles to a next-generation BS design that tightly couples integrated perception-communication-computation with an LAM-based cognitive core and concrete safety-critical scenarios. As summarised in Table~\ref{table_cmp}, representative works from these three lines and the proposed IBSA are compared along four dimensions: (i) functional focus of the next-generation BS (communication, perception, computation); (ii) type of BS-side core intelligence (LAM, LLM\footnote{The difference between LAM and LLM is detailed in \cite{chen2025wirelessNativeBigAI}.}, or no unified large model); (iii) whether the BS only passively receives wireless signals or can actively interact with and shape the environment via wireless actions; and (iv) targeted key scenarios (autonomous driving, low-altitude security). Table~\ref{table_cmp} highlights that existing architectures typically cover only a subset of these dimensions, whereas IBSA jointly realises all four and thus provides a concrete instantiation of a large-model-enabled embodied intelligent BS for integrated perception-communication-computation 6G systems.

\section{Architecture of Intelligent Base Station Agent}\label{sec_archi}
\begin{figure*}[!ht]
	\centering	
	\includegraphics[width=6in]{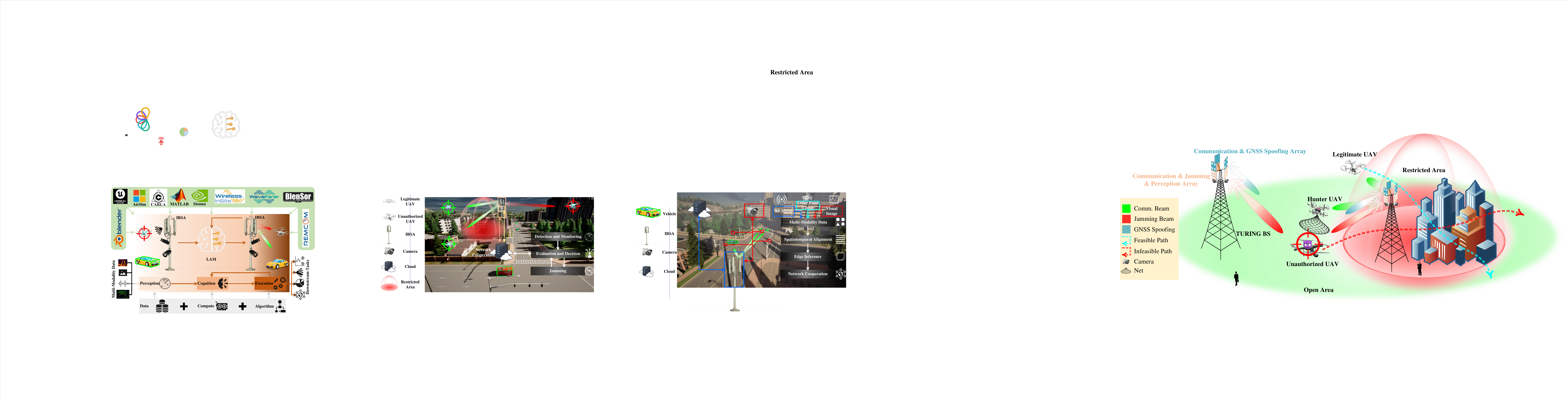}
	\caption{Architecture of {\color{black}embodied} intelligent BS agent.
	{\color{black}The framework is centered around an LAM that empowers the {\color{black}embodied} IBSA through a three-layer architecture, consisting of a perception layer, a cognition layer, and an execution layer. At the top, various simulation platforms are used to synthesize and replay scene data for closed-loop evaluation. The agent processes multi-modality data inputs, such as infrared, visual images, wireless signals, and lidar point clouds. This enables the embodied IBSA to perform various downstream tasks, which include prediction, beamforming, resource allocation, and network cooperation. The entire architecture is supported by the foundational components of data, compute resources, and algorithms.}}
	\label{fig_architecture}
\end{figure*}
\subsection{Overall Architecture and Deployment Forms of Embodied IBSA}\label{sec_archi_1}
With the evolution of 6G toward ``native intelligence" and ``perception-communication-computation integration," BSs cease to be mere passive forwarding nodes and become {\color{black}embodied} IBSAs endowed with {\color{black}perception}, cognition, and actuation capabilities.
The overall architecture of IBSA follows an upward three‐layer closed-loop:
\begin{itemize}
	\item Perception layer: acquires and pre‐processes multimodal data.
	\item Cognition layer: performs cross‐modal fusion, world modeling, and policy generation.
	\item Execution layer: carries out controllable actions on wireless resources and external entities.
\end{itemize}
Above these three layers, a {\color{black}LAM} provides a unified representation and inference interface, and is deployed across cloud-edge-end in a collaborative fashion to satisfy varying latency and energy constraints. As illustrated in Fig.~\ref{fig_architecture}, upper‐level multi‐engine simulation environments (e.g., Unreal Engine, Blender, AirSim, CARLA, MATLAB, Sionna, Wireless InSite, WaveFarer, Remcom, etc.) continuously feed multimodal data and digital‐twin feedback downward; multiple {\color{black}embodied} IBSA nodes in the mid‐tier complete perception-cognition-execution cycles under the drive of the LAM; and the lower tier’s data, compute resources, and algorithmic foundations underpin the entire agent ecosystem.

{\color{black}The perception layer, visually represented on the left in Fig.~\ref{fig_architecture} as ``Multi-Modality Data" flowing into the ``Perception" stage, emphasizes ``visible," ``audible," and ``measurable" perception.}
Inputs from the wireless domain include uplink/downlink pilots and channel state information (CSI), millimeter‐wave/terahertz (THz) radar echo matrices, and angle‑of‑arrival/time‑difference‑of‑arrival (AoA/TDoA) estimates; external sensors comprise roadside and vehicle‑mounted cameras, lidar, and thermal infrared imagers. Multimodal data must be filtered, compressed, and timestamped at the acquisition point and spatially registered to a common reference coordinate system. To control network load, event‑driven uploading, learned compression, or feature‑level upload strategies may be adopted. The perception layer also outputs data‑quality and uncertainty estimates, furnishing the cognition module with quantified confidence measures.

The cognition layer, {\color{black}which is fundamentally embodied by the LAM}, functions as the ``central brain" of {\color{black}embodied} IBSA, {\color{black}symbolized by the central LAM brain icon in Fig.~\ref{fig_architecture}}. At its core lies a multimodal large model with {\color{black}vision-language-radio coupling capability}, which maps heterogeneous signals into token sequences and captures intermodal mutual information and constraints via cross‑modal attention and spatiotemporal Transformers. Training draws on both real‐world network and road/airspace data (e.g., vehicle and UAV control logs) and long‑tail scenarios generated by digital‑twin platforms, employing self‑supervised or contrastive learning to bolster generalization and robustness. Mechanisms such as federated learning, parameter-efficient fine‑tuning (PEFT), and mixture-of-experts (MoE) enable model sharding across cloud, edge, and end devices, thus achieving privacy protection and low‑latency inference. The cognition layer should further incorporate memory and causal‑reasoning capabilities to support long‑term state tracking, threat‑intent analysis, and policy interpretability.

The execution layer, {\color{black}the final stage in the pipeline shown in Fig.~\ref{fig_architecture}}, implements cognitive outputs as controllable actions in both the communications and physical domains. On the wireless side, {\color{black}embodied} IBSAs may perform beamforming, power/spectrum scheduling, network‐slice allocation, and reconfigurable intelligent surface (RIS) configuration. On the security/control side, they can target unauthorized (``black‐fly") drones or anomalous vehicles with directional jamming, link spoofing, and coordinated behavior control (e.g., formation scheduling, obstacle‐avoidance path broadcasting), feeding execution outcomes back to the cognition layer to close the optimization loop~\cite{Qin2024}. 
Execution policies must balance multiple objectives, often within reinforcement‑learning or game‑theoretic frameworks and supplemented by safety constraints and human-machine collaboration interfaces to ensure controllability: communication quality of service (QoS) versus {\color{black}perception} accuracy, jamming efficiency versus collateral damage, automated decision‑making versus legal compliance.

From an engineering and deployment standpoint, {\color{black}embodied} IBSA can be classified by entity type, application domain, and deployment form. Fixed ground BSs are suited for continuous urban and highway coverage, whereas {\color{black}aerial BSs (UAV‑BS, balloon‑borne) provide rapid blind‑spot fill‑in during disasters or large events~\cite{ref_space_1,ref_space_2,ref_space_3}.} Application domains span V2X cooperative perception, UAV control, smart‑city security, and industrial internet-of-thing (IoT), each with distinct latency, reliability, and privacy requirements. Deployment follows a cloud-edge-end continuum: the cloud hosts ultra‑large models for retraining and global updates; the edge runs model shards or expert subnetworks for latency‑sensitive fusion inference and resource orchestration; end devices execute distilled or pruned lightweight models to enable degraded operation when disconnected. These tiers maintain consistency via parameter synchronization, knowledge distillation, and experience replay, while model routing and caching strategies alleviate network pressure.

The {\color{black}embodied} IBSA architecture, through its three-layer closed-loop of perception, cognition, and execution, tightly couples the triad of {\color{black} data, compute, and algorithms with the pathway of multimodal data, large models, and downstream tasks}.
Digital twins serve as data amplifiers for training and safe sandboxes for policy validation; large models provide cross‑task shared representations and few‑shot adaptation capabilities; and integrated communication/perception hardware with network orchestration ensures efficient real‑world realization of cognitive outcomes.

\subsection{Multimodal Large Model at the Cognition Layer}\label{sec_archi_2}
The cognition layer of an embodied IBSA is fundamentally embodied by a multimodal LAM that acts as the ``central brain". Beyond the high-level architecture described in Section~\ref{sec_archi_1}, we provide here a representative instantiation of such a model, consistent with recent multimodal sensing-and-communication datasets~\cite{ref_DeepSense, ref_DeepVerse, ref_Unreal, ref_M3SC, ref_SoM, yang2024v2x} and multimodal fusion architectures for autonomous driving and V2X~\cite{ref_MVX, ref_WC_Dataset, ref_HADAR, ref_24_arXiv_DeepSense_Predict, ref_TMLCN, ref_TN, ref_24_TITS_lidar_camera, ref_24_TIV_radar_camera, ref_23_ICRA_lidar_camera, ref_25_TITS_radar_camera, ref_24_IROS_radar_camera}.

1) Model structure: The multimodal LAM follows a Transformer-style encoder–decoder architecture with heterogeneous modality encoders and a shared spatiotemporal backbone. Wireless-domain inputs (uplink/downlink CSI, millimeter-wave (mmWave)/THz radar tensors, AoA/TDoA estimates) are first pre-processed into channel and radar feature tensors following the data formats and synchronization strategies adopted in multimodal sensing-and-communication datasets such as DeepSense 6G~\cite{ref_DeepSense}, Deepverse 6G~\cite{ref_DeepVerse}, M3SC~\cite{ref_M3SC}, SynthSoM~\cite{ref_SoM}, and V2X-radar~\cite{yang2024v2x}. These tensors are then fed into complex-valued convolutional or recurrent encoders to produce one-dimensional ``radio tokens".

Roadside and vehicle-mounted cameras are encoded by a vision Transformer (ViT)-style backbone into ``vision tokens", while lidar point clouds are processed by voxel-based or point-based encoders as in state-of-the-art sensor fusion frameworks~\cite{ref_24_TIV_radar_camera, ref_24_TITS_lidar_camera, ref_23_ICRA_lidar_camera, ref_25_TITS_radar_camera, ref_24_IROS_radar_camera}. Each token is augmented with learnable modality-type, spatial, and temporal embeddings so that heterogeneous inputs share a unified token space.

The resulting token sequences from all modalities are concatenated and passed through a stack of spatiotemporal Transformer blocks. Within each block, self-attention operates over the full multimodal sequence, while cross-modal attention explicitly enhances interactions between radio, vision, and point-cloud tokens, following the paradigm of MVX-ViT~\cite{ref_MVX} and related fusion architectures such as BEVFusion, MSSF, and CR3DT~\cite{ref_23_ICRA_lidar_camera, ref_25_TITS_radar_camera, ref_24_IROS_radar_camera}. On top of the shared backbone, multiple task-specific heads are attached for 3D object detection and tracking, link-quality or beam-index prediction, jamming/beamforming policy generation, and trajectory forecasting, similar in spirit to recent vehicular multi-sensor fusion designs~\cite{ref_24_TIV_radar_camera, ref_24_TITS_lidar_camera, ref_23_ICRA_lidar_camera, ref_25_TITS_radar_camera, ref_24_IROS_radar_camera}.

To make the design compatible with practical edge deployment, the IBSA-side LAM is designed to contain on the order of $10^9$ trainable parameters, which is comparable to compact multimodal Transformer models used in real-time perception and communications~\cite{ref_LAM_3}. Larger cloud-resident teacher models in the $10^{11}$–$10^{12}$ parameter range can be used for pre-training and distillation, while edge-side models are obtained via model compression and parameter-efficient fine-tuning~\cite{jiang2025comprehensive}.

2) Training data sources and learning objectives: The multimodal LAM is trained on a mixture of real and synthetic data. On the real-data side, we leverage public multimodal datasets that provide synchronized camera, lidar, and radar measurements~\cite{yang2024v2x, ref_24_Bosch_radar, ref_24_TIV_radar_camera}, together with communication-oriented datasets such as DeepSense 6G and M3SC that align CSI or radar signals with visual and point-cloud observations~\cite{ref_DeepSense,ref_M3SC}. On the synthetic side, joint traffic–wireless digital twins based on tools such as Wireless InSite, Sionna, and WaveFarer~\cite{ref_DeepVerse, ref_WirelessInSite, ref_sionna, ref_WaveFarer} are used 
to generate long-tail safety-critical scenarios and rare channel conditions that are difficult to capture in the field.

Training combines self-supervised objectives (e.g., masked token modeling and cross-modal contrastive learning) with supervised detection, localization, and beam-prediction losses, following recent multimodal learning practices in autonomous driving and ISAC~\cite{ref_24_TIV_radar_camera,
ref_24_TITS_lidar_camera, ref_23_ICRA_lidar_camera, ref_25_TITS_radar_camera, ref_24_IROS_radar_camera}. Federated learning, parameter-efficient fine-tuning, and MoE strategies~\cite{jiang2025large} further enable model partitioning across cloud–edge–end, providing privacy preservation and low-latency inference.

3) Integration of heterogeneous modalities: To integrate heterogeneous modalities in a principled manner, the LAM first performs spatiotemporal alignment at the feature level: raw CSI, images, and point clouds are time-stamped and projected into a common world coordinate frame, as recommended in recent surveys on multimodal fusion~\cite{ref_24_TIV_radar_camera,ref_24_TITS_lidar_camera, ref_23_ICRA_lidar_camera, ref_25_TITS_radar_camera, ref_24_IROS_radar_camera}. After tokenization, the model interleaves radio, vision, and point-cloud tokens within the same sequence and lets the Transformer attend over the full set. Cross-modal attention allows, for example, occluded vehicles that are only visible in radio-domain CSI to be associated with partially visible objects in the image space, while temporal attention stabilizes decisions across successive frames. In this way, the cognition layer naturally fuses ``through-the-obstacle" RF cues with high-semantic visual and geometric information, and propagates the fused representation to downstream communication and control heads.

\section{Key Application Scenarios}\label{sec_scenario}
Building on the general framework outlined above, we now turn to two illustrative application scenarios.
The first demonstrates how multi‑source fusion perception empowers robust, all-weather autonomous driving in complex traffic environments.
The second shows how ubiquitous {\color{black}embodied} IBSAs deliver full‑band, all‑modal safety monitoring and response in low‑altitude UAV operations.

\subsection{Cooperative Vehicle-Road Multi-Source Fusion Perception Empowering Autonomous Driving}\label{sec_auto}
\begin{figure*}[!ht]
	\centering	
	\includegraphics[width=6in]{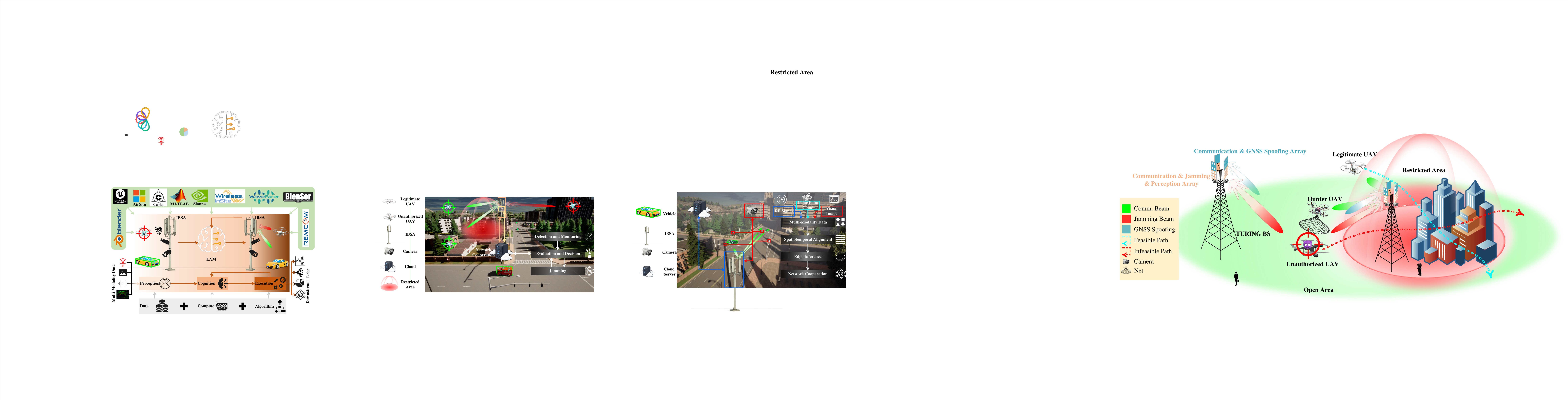}
	\caption{Cooperative vehicle-road multi-source fusion perception empowering autonomous driving.
	{\color{black}
	This figure illustrates the complete closed-loop workflow of multi-source fusion perception across the scenario, data, and network layers. The entities on the left, ``Vehicle, IBSA, Camera, and Cloud Server", delineate the primary roles in the system. Within the road scenario, the colored arrows represent data flows: blue arrows denote the uplink of environmental information from the embodied IBSA to the cloud; red arrows signify the acquisition and transmission of visual data from cameras; and green arrows indicate robust beamforming and high-fidelity communication from the embodied IBSA to vehicles, guided by perception results and cloud directives. The workflow on the right begins with the collection of ``Multi-Modality Data", including RF signals, visual images, and lidar point clouds. This data then proceeds through a vertical pipeline of ``Spatiotemporal Alignment", ``Edge Inference", and ``Network Cooperation" to achieve perception-communication-computation co-optimization for autonomous driving tasks. The overall diagram highlights the embodied IBSA's pivotal role as a perception-communication bridge within this closed-loop system.}
	}
	\label{fig_autonomous}
\end{figure*}

In urban canyons, highway on‑ramps, tunnel entrances, and traffic environments dense with large vehicles, single‑modality perception exhibits structural limitations: millimeter‑wave/THz links, owing to their strong directivity, are highly sensitive to blockage (when vehicles, pedestrians, or buildings intervene, link quality and detection capability degrade sharply)~\cite{ref_Zhang2025}; cameras, although offering high spatial resolution and rich semantic content, are acutely vulnerable to illumination changes and adverse weather, with nighttime glare, backlighting, and rain‑ or fog‑induced scattering frequently causing missed or false detections; lidar and conventional radar are respectively constrained by cost and angular resolution. Hence, the development of multi‑source information fusion mechanisms via vehicle-road-base‑station cooperation is widely recognized as the key to achieving all‑weather, all‑element safety in autonomous driving \cite{ref_MVX}.

Against this backdrop, the complementarity between cellular BSs and vision/point‑cloud sensors becomes increasingly pronounced.
BSs, operating in the sub‑6~GHz, {\color{black}upper-6~GHz centimeter-wave}, millimeter‑wave, and even THz bands, benefit from stable power supplies, wide‑area coverage, and deployable edge computing resources; by exploiting physical‑layer features such as channel state information (CSI) and beam‑training feedback signals, they can achieve non‑line‑of‑sight ``through‑the‑obstacle" perception. Roadside and vehicle‑mounted cameras and lidar, in turn, furnish high semantic density and fine spatial detail over local regions. These modalities complement one another in spatial scale (wide‑area vs. local), information dimension (physical channel vs. visual semantics), and robustness (occlusion vs. illumination): the BS can provide a priori target regions and velocity estimates within visual blind spots, while the visual modality feeds back to the communication side to refine beam selection, interference management, and resource scheduling, thereby realizing a co‑optimization of ``visibility" and ``connectivity" \cite{ref_MVX,ref_VTCFall_carla}.

To unlock this complementarity, {\color{black}the LAM's inherent capabilities for cross-modal alignment and shared representation learning are central.} Models such as MVX‑ViT (Multimodal/V2X vision Transformer) in~\cite{ref_MVX} encode wireless features into ``radio tokens" and image or point‑cloud features into ``vision tokens," then employ cross‑modal attention across spatial and temporal dimensions to establish robust associations even under occlusion or adverse illumination \cite{ref_MVX}. These architectures typically include multi‑task output heads that jointly predict object detection/tracking results and key communication metrics (e.g., optimal beam indices, link quality), thereby embedding perception feedback directly into network resource scheduling. 
Furthermore, by leveraging vision-radio consistency and temporal coherence constraints, self‑supervised or contrastive learning strategies enable cross‑modal alignment with minimal large‑scale annotations, substantially reducing labeling costs while enhancing generalization \cite{ref_24_arXiv_DeepSense_Predict}.

As illustrated in Fig.~\ref{fig_autonomous}, the entire multi‑source fusion workflow forms a coherent closed-loop across the scenario layer, data layer, and compute/network layer.
On the left, four abstracted entities, vehicle, {\color{black}embodied} IBSA, camera, and cloud server, delineate the roles of the in‑vehicle unit, {\color{black}embodied} IBSA, roadside camera, and cloud/edge server. In a real‑road scenario, blue arrows denote the uplink of environment information perceived by the {\color{black}embodied} IBSA; green arrows indicate robust beamforming and high‑fidelity data communication performed by the {\color{black}embodied} IBSA based on perceived vehicle targets and cloud directives; red arrows mark the acquisition and transmission of visual image streams. ``Multi‑Modality Data" comprises visual or infrared images from roadside cameras, wireless signals received by the {\color{black}embodied} IBSA, and lidar point clouds. The system then advances vertically through ``Spatiotemporal Alignment $\rightarrow$ Edge Inference $\rightarrow$ Network Cooperation": first realizing cross‑device and cross‑modal alignment in space and time; next executing fusion inference at the edge (e.g., via a lightweight MVX‑ViT or MoE expert network); and finally distributing results through V2X interfaces to achieve perception-communication-computation co‑optimization for autonomous driving tasks. The color coding and hierarchical framework visually convey modality complementarity, spatiotemporal synchronization, and the causal linkage between edge intelligence and network cooperation, while underscoring the {\color{black}embodied} IBSA's pivotal role as a perception-communication bridge within the closed-loop system.

To validate the effectiveness of this architecture, performance evaluation must span perception, communication, and agent dimensions. Perception metrics include mean average precision (mAP), recall rate, distance and velocity estimation errors, and cross‑modal matching accuracy; communication metrics encompass end‑to‑end latency (covering acquisition, synchronization, inference, and result dissemination), top-$k$ beam selection hit rate, link throughput and reliability, as well as uplink perception data volume and downlink broadcast frequency (network load); from the agent’s perspective, key indicators include task completion rate, collision‑rate reduction, model parameter count and energy consumption, and interpretability. A unified key performance indicator (KPI) framework is essential to quantify overall benefits and to guide decisions on model compression, compute scheduling, and bandwidth allocation \cite{ref_TN}.

Existing research can be categorized into three main threads. First, multimodal cooperative perception datasets and benchmarks (e.g., MVX‑Dataset, MultiX~\cite{ref_MultiX}) provide annotation and evaluation frameworks for joint vision-radio tasks but predominantly focus on vehicle-to-infrastructure (V2I) scenarios and lack large-scale public base-station-level wireless raw data~\cite{ref_Li2024}.
Second, digital-twin-driven simulation platforms are maturing: by coupling CARLA with wireless ray-tracing tools such as Sionna or Wireless InSite, they enable real-to-twin close-loop optimization; representative studies {\color{black}\cite{ref_VTCFall_carla,ref_TN}} explore beam selection and latency-accuracy trade‑offs. Third, large‑model architectures, including MVX‑ViT and radio-vision Transformer, demonstrate significant potential but remain challenged by parameter scale, inference latency, and edge‑deployment constraints, leaving large‑scale real‑world deployment still out of reach \cite{ref_MVX}. Overall, unified interfaces, standardized evaluation protocols, and real‑road validation remain underdeveloped in the public domain.

Despite these advances, multiple challenges persist in data privacy and security, standardized interfaces, and real‑world deployment trials. Vehicle and roadside video often contain sensitive personal information, necessitating advanced privacy-preserving mechanisms far beyond basic data masking. While federated learning and differential privacy provide a baseline by keeping data local and adding noise, they may not sufficiently protect against inference attacks on shared model gradients.
To address this, robust anonymization schemes are required. Homomorphic encryption is a powerful tool, allowing the IBSA or cloud server to perform computations (e.g., model aggregation) directly on encrypted data without decryption, thus ensuring sensitive trajectories and features remain confidential even from the coordinator~\cite{tianTVT}. Further enhancements include integrating federated learning with secure multi-party computation to distribute the aggregation task, or using zero-knowledge proofs to allow vehicles to prove they possess valid data without revealing the data itself~\cite{yuca2025survey}.

Wireless data likewise may reveal trajectories and identities, requiring a robust combination of these cryptographic, policy, and technical safeguards\cite{ref_24_arXiv_DeepSense_Predict}. Concurrently, the absence of industry standards for cross‑modal data formats, synchronization protocols, and model APIs impedes multi‑vendor collaboration and large‑scale adoption. Transitioning from laboratory to roadway demands system‑level trade‑offs among fifth-generation (5G)/6G network slicing, edge compute scheduling, device heterogeneity, and cost, alongside the establishment of open testbeds and sandbox policies \cite{ref_TN}. Furthermore, to address dynamic variations in construction, seasons, and traffic flow, models must support continual learning and domain adaptation; under constrained bandwidth and compute resources, joint optimization of sampling rates, upload strategies, inference accuracy, and scheduling policies will be the core challenge for the long‑term evolution of multi‑source fusion-enabled autonomous driving \cite{ref_VTCFall_carla,ref_24_arXiv_DeepSense_Predict}.

\subsection{Ubiquitous {\color{black}Intelligent Base Stations Agent} Empower Low-Altitude Safety}\label{sec_LAE}
\begin{figure*}[!ht]
	\centering	
	\includegraphics[width=6in]{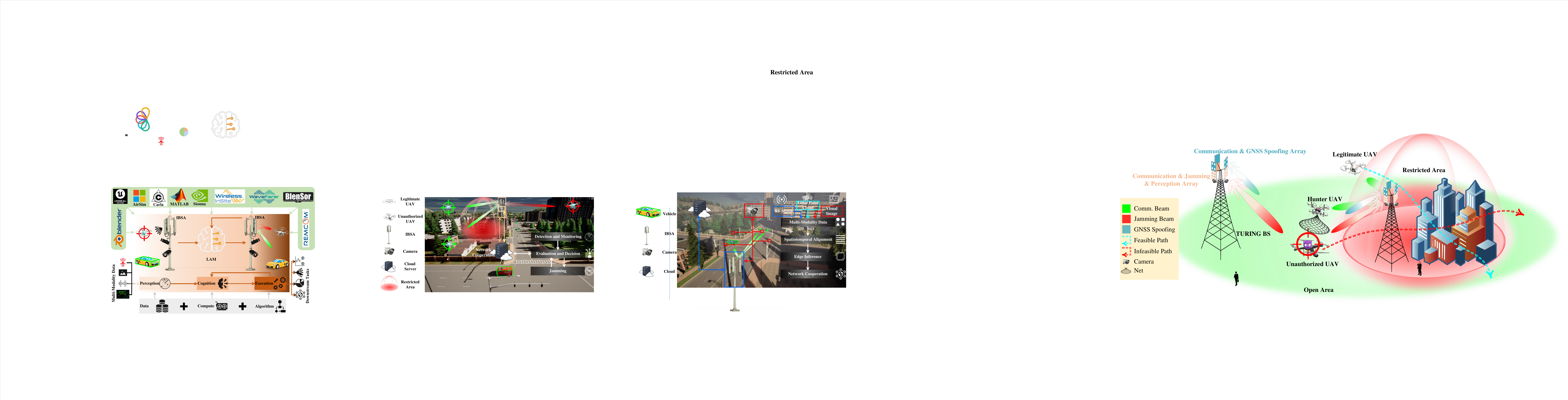}
	\caption{Ubiquitous {\color{black}embodied} IBSA empower low-altitude safety.
	{\color{black}
	This figure illustrates the closed-loop process for low-altitude security enabled by the IBSA, detailing entity roles, information flows, and the decision chain. In the central scenario, green arrows represent the continuous monitoring and data uplink for legitimate UAVs, while red arrows signify the early warning, targeting, and eventual jamming of unauthorized or suspicious UAVs entering a restricted area. The vertical arrows indicate data uploads from embodied IBSAs and cameras to the edge/cloud server, and the white bidirectional arrows depict the crucial role of ``Network Cooperation" among multiple embodied IBSAs and cloud-edge nodes for tasks like collaborative localization and resource scheduling. The workflow on the right outlines the full decision chain, progressing from ``Detection and Monitoring", through ``Evaluation and Decision", to responsive actions like ``Jamming" or issuing control commands to neutralize threats. This schematic underscores the embodied IBSA's central function in an integrated system that provides perception, assessment, and interference capabilities for low-altitude airspace management.
	}
	}
	\label{fig_LAE}
\end{figure*}
Low‑altitude airspace is rapidly transitioning from sparse control to dense openness, and the widespread use of UAVs in logistics, inspection, and emergency response brings real threats of unauthorized flights and airspace incursions. A typical adversarial scenario involves small, uncrewed rotorcraft loitering over urban cores, energy hubs, or large‐scale events; their low radar cross‑section, agile flight paths, and sudden appearances evade traditional single‑modality monitoring. Any misclassification or missed detection incurs substantial legal liability and social cost.
Consequently, anchored by {\color{black}embodied} IBSA with their {\color{black}wide‑area coverage and edge computing advantages}, constructing an all‑weather, full‑band, multi‑modality low‑altitude security perception and response closed-loop has become a critical component of the ``new air traffic management" system.

At the heart of this objective lies the fusion of BS side wireless {\color{black}perception} with heterogeneous modalities such as vision and acoustics. On the wireless side, uplink/downlink pilots, CSI anomaly patterns, millimeter‑wave/THz radar echoes, and AoA estimates enable non-line‑of‑sight detection of high‑speed or occluded targets. Vision and thermal‑infrared cameras supply fine‐grained shape information and rotor‐blade spectral cues, while microphone arrays capture airframe‑specific acoustic signatures. Multi‑modality collaboration maintains high detection confidence even when a single modality fails (e.g., at night, in dense fog, or under strong interference), and mutual corroboration suppresses false alarms. Datasets such as DeepSense 6G in~\cite{ref_DeepSense} demonstrate the importance of synchronized acquisition and temporal alignment for multi‑modal communications-perception paradigms, offering reusable templates for data‑structure design and unified evaluation metrics; Great-X/Great-MSD single-engine simulator in~\cite{ref_Unreal} realizes synchronous CSI, RGB, and lidar generation in low‑altitude UAV scenarios, validating CSI‑driven 3D localization and cross‑engine generalization, and providing high‑fidelity support for ``real-twin" co‑modeling and algorithm verification.

As depicted in Fig.~\ref{fig_LAE}, this subsection's schematic illustrates the {\color{black}embodied} IBSA‑enabled low‑altitude security closed-loop across three dimensions: entity roles, information flows, and decision chains. At the scene center, green arrows denote continuous monitoring and data uplink for authorized UAVs, while red arrows indicate early warning and targeting of suspicious or intruding UAVs. Orange and yellow vertical arrows mark data uploads from {\color{black}embodied} IBSA/cameras to edge servers, and white bidirectional arrows show collaboration among multiple base‑stations and cloud‑edge nodes. On the right, the ``Detection and Monitoring $\rightarrow$ Evaluation and Decision $\rightarrow$ Jamming" pipeline captures the full chain from perception through assessment to interference or hijacking. The diagram underscores the pivotal role of the ``Network Cooperation" module in multi‑point collaborative localization, interference resource scheduling, and strategy dissemination~\cite{ref_Xia2024}.

At the algorithmic level, {\color{black}LAMs} furnish unified representations for UAV identification, trajectory prediction, and threat assessment~\cite{Zhang2025}. 
Heterogeneous inputs, such as wireless CSI sequences, visual frames, acoustic spectrograms, are tokenized into a common sequence, and cross‑modal attention with temporal modeling performs identity classification and intent inference.
Generative or world models (e.g., Great‑X/Great‑MSD, DeepSense's digital‑twin scenarios) can synthesize ``long‑tail threat trajectories" and ``complex interference environments" within simulation domains, enabling risk‑adaptive evaluation and game‑theoretic strategy optimization. To meet continual learning demands, federated learning, PEFT, and incremental contrastive learning support online model updates and domain generalization without uploading raw sensitive data.

Designing countermeasures requires balancing effectiveness, controllability, and compliance. For jamming, reconfigurable intelligent surfaces (RIS) or coordinated multi‑BS beamforming can apply directional power suppression or link spoofing to a target UAV, avoiding indiscriminate electromagnetic emissions and collateral damage. For hijacking or forced landing, precise localization confirmed by vision/acoustics enables low‑power global position system (GPS) or command‑link spoofing to guide the UAV into a safe landing zone. Localization and tracking leverage multi‑BS TDoA and AoA fusion with particle filtering to reconstruct trajectories at centimeter‑level accuracy, supporting forensic evidence and post‑operation analysis.
It must be stressed that such active interference is not a standard network function. Under international frameworks such as the international telecommunication union (ITU) radio regulations~\cite{itu2020radio}, intentional jamming is broadly prohibited to prevent harmful interference with legitimate radiocommunication services.
It is imperative that any active interference therefore operates as a specific, national-level legal exception, strictly under emergency authorization (e.g., by public safety or defense authorities) and within designated frequency bands, as discussed below.

Regarding experimental benchmarks and public datasets, real low‑altitude multi‑modal data remain scarce. To date, only Great-MSD in a single‑engine environment has constructed large‑scale synchronized UAV trajectory, CSI, and vision datasets, providing benchmark experiments for CSI‑driven 3D localization. DeepSense 6G offers limited UAV scenarios but primarily focuses on V2I communications and ground‑based perception. There is an urgent need for standardized low‑altitude security data formats (including threat labels, interference strategies, and legal statuses), unified evaluation metrics ({\color{black}detection rate, false alarm rate, and response latency triplets}), and cross-domain transfer and adversarial‐robustness assessment protocols.

Although multi-station architectures exhibit significant potential, they face multiple challenges. First, legal compliance and liability allocation: UAV detection, jamming, and forced landing involve air-traffic control, public security, and radio-management authorities. As noted, the act of jamming is generally illegal under ITU regulations~\cite{itu2020radio}. Therefore, the ``authorization" process mentioned is a critical legal procedure, typically reserved for designated national agencies (e.g., ministry of defense, national police) to counter immediate security threats. This capability cannot be deployed by civil or commercial IBSA operators without such explicit, lawful delegation~\cite{kutynska2023legal}. Balancing this emergency response authority with privacy protection demands a technology-policy co-design.
Second, false‑alarm costs and trust calibration: misclassifying a legitimate UAV as a threat entails economic and legal risks, necessitating confidence thresholds, self‑auditing, and human-in-the-loop review mechanisms. Third, collaborative asynchronous transfer mode and standard interfaces: the lack of cross‑operator and cross‑vendor data interfaces and synchronization protocols limits multi‑BS cooperation and cross‑domain coordination. Fourth, real‑world deployment trials: edge‑compute overhead, network slicing orchestration, device heterogeneity, and cost control require phased validation from simulation through testbeds to live urban environments. Looking ahead, establishing open low‑altitude safety sandboxes, advancing privacy‑preserving multimodal data computation, and developing joint communication-perception-control resource orchestration algorithms will be the key pathways for ``ubiquitous BSs" to ensure low‑altitude security.

\subsection{Generalizability Analysis of Heterogeneous IoT Scenarios}
While the preceding subsections focus on autonomous driving and low-altitude safety, the proposed LAM-enabled IBSA architecture possesses inherent generalization capabilities that extend to other critical 6G scenarios, most notably industrial IoT and smart cities. The core ``perception-cognition-execution" closed-loop, driven by the multimodal fusion capabilities of LAMs, serves as a universal solution for heterogeneous environments characterized by visual occlusion and complex dynamics.
\subsubsection{Industrial IoT} 
The smart factory represents an extreme environment that challenges traditional perception systems. Unlike open roads, factories are dense with metallic racks, moving robotic arms, and high-density equipment, creating severe visual occlusion and complex multipath wireless propagation. Autonomous mobile robots (AMRs) often face blind spots that onboard perception devices (e.g., lidar) cannot resolve. In this context, the IBSA functions as critical infrastructure for the factory, generalizing the ``vehicle-road cooperation" model (Section~\ref{sec_auto}) into ``AMR-infrastructure collaboration". Leveraging the ``through-the-obstacle" nature of wireless signals, the IBSA fuses data from surveillance cameras with uplink CSI and millimeter-wave radar perception. This allows the system to perceive AMRs or humans hidden behind metallic obstructions, where visual perception fails.
Furthermore, the edge-deployed LAM processes this multimodal stream to construct a real-time, dynamic ``world model" of the factory floor, predicting the trajectories of both humans and machines. Replicating the safety loop of low-altitude scenarios (Section~\ref{sec_LAE}), the IBSA ensures human-machine safety via its execution layer. Upon predicting a potential collision, the system does not merely alert but triggers millisecond-level commands (e.g., emergency braking for robotic arms) via ultra-reliable low-latency communication, realizing a true integration of perception, communication, and computation.

\subsubsection{Smart City}
At a macroscopic level, the IBSA architecture scales to support smart city applications. Here, the IBSA acts as a ubiquitous intelligent node, an ``urban neuron"—managing diverse data streams ranging from traffic flows to smart grid loads. The traffic management optimization (e.g., ``Green Wave" for emergency vehicles) or intelligent load balancing strategies in smart grids follows the same ``perception-cognition-execution" paradigm found in mobility scenarios. The LAM at the edge serves as a regional brain, fusing heterogeneous data (visual traffic data, power consumption metrics) to generate optimal control policies for resource allocation. 

In summary, the applicability of the IBSA is not limited to specific verticals. Its ability to convert multimodal perception data into intelligent decisions and close the loop through wireless execution proves its potential as a generalized infrastructure for the heterogeneous scenarios of the 6G era.

\section{Key Enabling Technologies of Embodied IBSA}\label{sec_enablingTech}
In the previous section, we established the overall architecture of {\color{black}embodied} IBSA, clarifying its large-model-centric design, the closed ``perception-cognition-execution" loop, and its reliance on cloud-edge-end collaboration. However, to realize this architecture in practice and ensure stable operation in complex, dynamic environments, a suite of key enabling technologies must work in concert across computation, data, algorithms, security, and governance. 

Among these, the LAM itself forms the cognitive heart of the IBSA. To detail this core component, Table~\ref{tab:key_technologies} provides a specific summary of the key machine learning techniques that empower this ``central brain," from its foundational architecture to its training and inference methods~\cite{ref_LAM_1, ref_LAM_3, ref_Embodied_2, yang2025generative, zhou2025perception, chen2025wirelessNativeBigAI,IoTJMFengZhiyong}. 
	
Complementing this, Table~\ref{tab:section4_summary} provides a high-level overview of the key system-level technologies discussed in this section, outlining their respective roles from perception and simulation to resource orchestration and security. The following subsections will now elaborate on these critical topics in detail.

\begin{table*}[!t]
	\centering
	\caption{Key LAM technologies for IBSA.}
	\label{tab:key_technologies}
	\renewcommand{\arraystretch}{1.3} 
	\small 
	\begin{tabularx}{\textwidth}{@{} l l X @{}}
		\toprule
		\textbf{Category} & \textbf{Key Technology} & \textbf{Introduction \& Purpose} \\
		\midrule
		
		\multirow{6.75}{*}{\textbf{Infrastructure}} 
		& Transformer & Cornerstone of LLMs. Uses self-attention for parallel processing and long-range dependency capture. \textbf{Purpose:} Processing 6G time-series signals (e.g., CSI, radar waveforms) and modeling complex spatiotemporal correlations. \\
		& \multirow{2}{*}{\shortstack[l]{SSM (State Space  \\Model) / Mamba}}   & Efficient alternative to Transformer with linear complexity. \textbf{Purpose:} Provides faster inference and lower memory usage for ultra-long wireless signal sequences, meeting real-time requirements. \\
		& Multimodal Fusion & Maps vision, text, and sensor data to a unified semantic space (e.g., CLIP~\cite{yang2025generative}/ViT). \textbf{Purpose:} Theoretical basis for the IBSA perception layer, enabling heterogeneous fusion of RF signals and visual images. \\
		\midrule
		
		\multirow{10}{*}{\textbf{\shortstack[l]{Training \&\\Alignment}}} 
		& Instruction Tuning & Fine-tuning via high-quality instruction data. \textbf{Purpose:} Enables IBSA to understand and execute natural language control instructions from network management (e.g., ``optimize beams'', ``activate jamming''). \\
		& \multirow{2}{*}{\shortstack[l]{RLHF (Reinforcement Learning\\from Human Feedback)}}  & Optimizes policy via reward models to align with human values. \textbf{Purpose:} Constrains execution layer decisions to strictly follow safety rules and interference limits while maximizing throughput. \\
		& PEFT / LoRA & Freezes model body, training few parameters. \textbf{Purpose:} Lowers compute barriers, enabling low-cost, rapid adaptation at edge BSs for specific environments without full retraining. \\
		& Federated Learning & Distributed privacy-preserving training uploading only gradients. \textbf{Purpose:} Solves data silos and enables collaborative training while protecting user privacy (e.g., vehicle trajectories). \\
		\midrule
		
		\multirow{6.5}{*}{\textbf{\shortstack[l]{Inference \&\\Acceleration}}} 
		& MoE & Splits model into expert sub-networks with sparse activation. \textbf{Purpose:} Dynamically invokes modules based on tasks (comm. vs. perception) to reduce latency and energy on resource-constrained BSs. \\
		& FlashAttention & Input/Output-aware exact attention acceleration. \textbf{Purpose:} Boosts GPU inference speed, crucial for ms-level response scenarios like V2X collaborative perception. \\
		& Quantization & Reduces numerical precision (e.g., INT4) to compress models. \textbf{Purpose:} Reduces memory footprint, essential for deploying large models from cloud to edge BSs. \\
		\midrule
		
		\multirow{5}{*}{\textbf{\shortstack[l]{Agents \&\\Decision}}} 
		& In-Context Learning & Few-shot learning without parameter updates. \textbf{Purpose:} Enables rapid adaptation to unseen channel environments or new interference patterns via input examples. \\
		& Chain of Thought & Guides step-by-step reasoning. \textbf{Purpose:} Improves accuracy and interpretability in complex logic tasks (e.g., drone intent analysis, fault diagnosis). \\
		& Embodied Agent & Closed-loop interaction with physical environments. \textbf{Purpose:} Core IBSA concept; empowers BSs to perceive, plan, and execute physical actions (beam adjustment, jamming). \\
		\midrule
		
		\multirow{2}{*}{\textbf{\shortstack[l]{Data \&\\Generation}}}  
		& \multirow{2}{*}{World Models} & Digital twins and simulation of physical environments. \textbf{Purpose:} Generates synthetic data obeying physical laws (e.g., collisions) to address scarcity of extreme scenario data. \\
		
		\bottomrule
	\end{tabularx}
\end{table*}
\begin{table*}[!t]
	\centering
	\caption{Overview of key enabling technologies for the embodied IBSA system.}
	\label{tab:section4_summary}
	\renewcommand{\arraystretch}{1.4} 
	\small 
	\begin{tabularx}{\textwidth}{@{} l >{\raggedright\arraybackslash}p{4.5cm} X @{}}
		\toprule
		\textbf{Category} & \textbf{Key Technology} & \textbf{Role and Significance for Embodied IBSA} \\
		\midrule
		
		\multirow{8.5}{*}{\textbf{\shortstack[l]{Cognitive Core\\ \textit{(Perception \&} \\ \textit{Reasoning)}}}} 
		& \textbf{Advanced Perception \& Fusion} (Sec.~\ref{sec_AdvancedPerception}) & Fuses external sensor data (camera, lidar) with internal wireless signals to overcome individual sensor limitations (e.g., occlusion, weather). Edge deployment ensures millisecond-level inference for safety-critical tasks. \\
		\cmidrule(l){2-3} 
		& \textbf{Edge LAM Training \& Inference} (Sec.~\ref{sec_edgeLAMtrain}) & Manages the LAM's cognitive core on resource-constrained edge devices. Uses PEFT (LoRA) for rapid, low-memory adaptation and MoE to reduce latency via ``compute-on-demand". \\
		\cmidrule(l){2-3}
		& \textbf{Interpretability} (Sec.~\ref{sec_interpret}) & Moves beyond standard methods (SHAP) to explain complex spatiotemporal and multimodal fusion decisions. Provides verifiable causal links (e.g., ``why this beam was chosen") for regulatory compliance and operator trust. \\
		\midrule
		
		\multirow{7.5}{*}{\textbf{\shortstack[l]{System Orchestration\\ \textit{(Architecture)}}}}
		& \textbf{Cloud-Edge-End Coordination} (Sec.~\ref{sec_cloudedgeend}) & Implements the LAM across heterogeneous hardware by deploying full models in the cloud, medium models at the edge, and lightweight heads on end devices. Uses model compression (quantization, pruning) and gradient compression for efficient updates. \\
		\cmidrule(l){2-3}
		& \textbf{Resource Co-optimization} (Sec.~\ref{sec_resource}) & Manages the fundamental trade-off between Perception, Communication, and Computatio. Uses network slicing to dynamically allocate resources (GPU, spectrum) to ensure QoS for diverse tasks (e.g., high-priority safety vs. low-priority comms). \\
		\midrule
		
		\multirow{5.5}{*}{\textbf{\shortstack[l]{Security \& Governance\\ \textit{(Privacy \& Trust)}}}}
		& \textbf{Federated Learning \& Privacy} (Sec.~\ref{sec_FL}) & Enables collaborative IBSA model training across multiple BSs and vehicles without sharing raw, sensitive user data (e.g., trajectories). Enhanced with differential privacy or homomorphic encryption for robust security. \\
		\cmidrule(l){2-3}
		& \textbf{Security \& Trustworthiness} (Sec.~\ref{sec_securitytrust}) & Defends the IBSA against multi-dimensional attacks, including adversarial examples (in vision or RF), model inversion, and backdoors. Ensures accountability through auditable logs for decision traceability. \\
		\midrule
		
		\multirow{3.5}{*}{\textbf{\shortstack[l]{Foundation\\ \textit{(Data Engine)}}}} 
		& \textbf{Digital Twin \& Simulation} (Sec.~\ref{sec_digitaltwin}) & Provides a safe, controllable, and reproducible environment for training and validating the IBSA's perception-cognition-execution loop. Generates synchronized multimodal data (RF, vision, lidar) and bridges the ``sim-to-real" gap. \\
		
		\bottomrule
	\end{tabularx}
\end{table*}
\subsection{Cognitive Core: Algorithms for Perception and Reasoning}
\subsubsection{Advanced Perception and Fusion Capabilities}\label{sec_AdvancedPerception}
To achieve robust environmental perception, as required by the first layer of the architecture in Fig.~\ref{fig_architecture}, the embodied IBSA overcomes the limitations of individual sensors. It fuses high-resolution data from external cameras and lidar with its own unique ability to sense using communication signals, which can penetrate obstacles and adverse weather that impair visual systems.
This fusion process represents the critical first stage of the perception–cognition–execution optimization loop, where the choice of data modality and fusion complexity directly impacts the subsequent computational load (cognition) and the timeliness of responsive actions (execution).
To make this fused data actionable for safety-critical tasks, all computation is performed at the network edge, enabling millisecond-level inference and decision-making.
This edge deployment is essential for managing the inherent trade-offs between perception accuracy and latency within the closed-loop.
Moreover, the system’s reliability in unpredictable, ``long-tail'' scenarios is continuously hardened by training on vast datasets generated within digital twin environments and by using self-supervised learning to enhance generalization.

\subsubsection{Edge LAM Training and Inference Optimization}\label{sec_edgeLAMtrain}
{\color{black}Embodied} IBSA's large‑model core must perform low‑latency inference and on‑demand fine‑tuning at the edge. Parameter‑efficient fine‑tuning methods such as low-rank adaptation (LoRA) train only low‑rank adapter matrices while keeping the backbone weights frozen, enabling rapid domain adaptation and significantly reducing memory footprint and communication overhead. The {\color{black}MoE} architecture sparsely activates expert subnetworks, mapping different wireless scenarios and task types {\color{black}(perception/communication/control)} to specific experts for ``compute‑on‑demand" allocation. To prevent routing oscillation and load imbalance, temperature-controlled top-$k$ gating and load‑balancing losses can be employed~\cite{ref_MoE}.
Hierarchical scheduling addresses heterogeneous cloud-edge-end resources.
Scheduling policies should jointly consider graphics processing unit (GPU) utilization, link conditions, and task urgency, and may leverage reinforcement learning or game‑theoretic optimization for bandwidth-computation co‑orchestration.

\subsubsection{Interpretability Considerations}\label{sec_interpret}
Since {\color{black}embodied} IBSA's decisions directly impact traffic safety and low‑altitude management, interpretability is a prerequisite rather than an afterthought. Beyond standard attention visualization and feature‑attribution methods {\color{black}(SHAP~\cite{ref_SHAP}, Integrated Gradients~\cite{ref_IG})}, structured explanations for spatiotemporal sequences and multimodal fusion are required: explicit wireless-vision alignment weights, expert selection paths, and beamforming adjustment causal chains. Furthermore, symbolic reasoning or causal graph models can produce verifiable intermediate conclusions, and a natural‑language explanation head can articulate decision rationales to operators, enhancing human-machine collaboration and regulatory transparency.

\subsection{System Architecture: Coordination and Orchestration}
\subsubsection{Cloud-Edge-End Coordination and Model Compression Strategies}\label{sec_cloudedgeend}
To accommodate heterogeneous computational and storage capabilities across nodes, model compression techniques, such as pruning, quantization, knowledge distillation, and tensor decomposition, must be combined.
End devices retain only task-specific heads or lightweight transformers, edge nodes maintain medium‑scale backbones with online update capability, and the cloud preserves full‑parameter models and version histories. Coordination mechanisms must address consistency (version conflicts, cache staleness), hardware heterogeneity, and link instability. {\color{black}Gradient compression (e.g.,\ top‑$k$~\cite{ref_topk}, random sparsification~\cite{ref_random_sparsification}) and joint encoding-scheduling design ensure timely synchronization under limited uplink bandwidth~\cite{ref_joint_encoding_scheduling}.}

\subsubsection{Resource Orchestration: Network Slicing and Perception-Communication-Computation Co-optimization}\label{sec_resource}
Embodied IBSA concurrently supports communication services, perception tasks, and intelligent computation, which forms a complex perception–cognition–execution collaborative optimization problem.
This requires network slicing to partition spectrum, time slots, cache, and GPU cores into logical resource pools for dynamic task orchestration under diverse QoS/quality of experience requirements. 
This co-optimization can be formulated as a joint-constraint multi-objective problem: minimize end-to-end latency and energy consumption while maximizing perception accuracy, cognitive inference quality, and communication throughput~\cite{Chen2024}.
Solutions may leverage distributed Lagrangian decomposition, deep reinforcement learning, or graph neural networks on large‑scale topologies to manage these complex trade-offs~\cite{jiang2025comprehensive,ref_LAM_3}. For bursty tasks (e.g.,\ UAV intrusion), preemptive slice reconfiguration and priority scheduling must be supported to guarantee real‑time performance for critical missions.

\subsection{Security and Trustworthiness: Privacy and Defense}
\subsubsection{Federated Learning and Privacy Protection}\label{sec_FL}
When multiple BSs, vehicles, and cameras collaboratively update the {\color{black}distributed LAMs}, raw data are often constrained by privacy and bandwidth. federated learning enables nodes to upload only gradients or model deltas; combining adaptive aggregation schemes {\color{black}(e.g.\ FedAvgM~\cite{ref_FedAvgM}, FedProx~\cite{ref_FedProx})} mitigates non-i.i.d. data distribution issues. To further enhance privacy and robustness, techniques such as differential privacy noise, secure multi‑party computation, or homomorphic encryption can be {\color{black}introduced~\cite{ref_yang2019federated}}. Meanwhile, {\color{black}embodied} IBSA must defend against malicious clients and poisoning attacks; robust aggregation methods based on median or trimmed mean, reputation scoring, and anomaly detection mechanisms ensure trustworthy model evolution in {\color{black}open environments~\cite{ref_cao2020fltrust}}.

\subsubsection{Security and Trustworthiness: Adversarial Robustness, Model Evaluation, Accountability}\label{sec_securitytrust}
As the ``intelligent brain" of critical infrastructure, embodied IBSA must withstand multidimensional attacks: adversarial examples can compromise vision/RF recognition; model inversion may leak sensitive scene data; backdoors can trigger uncontrolled behavior under specific stimuli. 

For wireless-domain threats such as CSI tampering, defense strategies include channel-consistency verification across multiple antennas, physical-layer authentication using device-specific RF fingerprints, and anomaly detection based on statistical deviation from expected channel behavior~\cite{MENG2025104085}. For visual adversarial examples, certified defenses such as randomized smoothing, input transformation (e.g., JPEG compression, spatial smoothing), and ensemble-based detection can effectively mitigate perturbation-based attacks~\cite{carlini2023certified}. Adversarial training, which augments training data with projected gradient descent perturbations, remains a primary defense to improve model robustness against both modalities.

Regarding privacy protection, differential privacy can be integrated at multiple levels: local differential privacy adds calibrated Laplacian or Gaussian noise to raw sensor readings before upload; central differential privacy applies noise during gradient aggregation in federated learning; and the privacy budget should be carefully tuned to balance utility and protection~\cite{ref_yang2019federated}. 

For evaluation, comprehensive security benchmarks covering the full perception-cognition-execution chain are needed, encompassing robustness, transferability, and false positive/negative trade-offs. On accountability, auditable logs, decision-traceability graphs, and compliance interfaces delineate responsibilities between the model and operators to satisfy regulatory and legal requirements.

\subsection{Digital Twin and Simulation Platforms}\label{sec_digitaltwin}
{\color{black}Training and validating the IBSA's LAM requires rapid iteration under safe, controllable, and reproducible conditions, a need met by the digital twin and simulation platforms illustrated at the top of Fig.~\ref{fig_architecture}.} Photo-realistic traffic and vision simulators such as CARLA~\cite{ref_carla}, AirSim~\cite{ref_AirSim}, Blender~\cite{ref_blender}, together with wireless channel simulators like Sionna~\cite{ref_sionna}, Wireless InSite~\cite{ref_WirelessInSite}, WaveFarer~\cite{ref_WaveFarer}, REMCOM, form an end‑to‑end digital twin platform. By employing a unified scene‑description language and data bus, multimodal data (vision, lidar point clouds, radio-frequency signals, positioning) are collected synchronously; domain randomization and procedural generation help bridge the sim‑to‑real gap.
Achieving this cross-platform temporal alignment is a critical challenge, often addressed using hardware triggers (e.g., precision time protocol) or middleware like the robot operating system to apply unified timestamps to heterogeneous data streams, as demonstrated in practical dataset generation efforts~\cite{yang2024v2x}.
Furthermore, closed-loop simulation validates the perception-decision-action chain offline, and synthetic data feed back into large‑model pre‑training and sparse scenario augmentation.

These enabling technologies collectively underpin {\color{black}embodied} IBSA's transition from a ``large‑model system" to a ``deployable, governable, large‑model‑driven agent." They form a closed-loop across computation, data, algorithms, resource management, and security compliance, providing the technical foundation for realizing the integrated perception-communication-intelligence vision of 6G.

\begin{table*}[!t]
	\centering
	\caption{Holistic evaluation framework for embodied IBSA (summary of Section~\ref{sec_metrics}).}
	\label{tab:evaluation_framework}
	\renewcommand{\arraystretch}{1.4} 
	\small 
	\begin{tabularx}{\textwidth}{@{} >{\RaggedRight}p{3.2cm} >{\RaggedRight}X >{\RaggedRight}X @{}}
		\toprule
		\textbf{Evaluation Pillar} & \textbf{Key Metrics / Components} & \textbf{Purpose \& Description} \\
		\midrule
		
		\textbf{Perception Metrics} (Sec.~\ref{sec_perceptionmetrics}) 
		& FPR, FNR, Precision/Recall (F1), DET Curve, mAP, Localization Error (RMSE), 2D/3D IoU, Velocity/Angle Error, Missing-Rate.
		& To quantify the IBSA's accuracy, reliability, and precision in environmental perception and object detection. \\
		\hdashline
		
		\textbf{Network Metrics} (Sec.~\ref{sec_networkmetrics}) 
		& Throughput (Gbps/Hz), End-to-End Latency (ms, 99.9th percentile), Energy per Inference (J), Interference Level (SINR/BLER degradation).
		& To measure the communication efficiency, response speed (especially tail latency), resource consumption, and impact on existing network services. \\
		\hdashline
		
		\textbf{Agent-Level Metrics} (Sec.~\ref{sec_agentmetrics}) 
		& Task Completion Rate (TCR), Robustness Drop (under adverse conditions).
		& To evaluate the IBSA's overall effectiveness and resilience (robustness) in executing its complete "perception-cognition-execution" closed-loop tasks. \\
		\hdashline
		
		\textbf{Benchmark Resources} (Sec.~\ref{sec_benchmark}) 
		& \textbf{Public Datasets:} Provide annotated real-world samples (e.g., vision, radar, wireless channels).
		\textbf{Digital-Twin Platforms:} (e.g., CARLA + Sionna) Generate controllable, synchronized synthetic data and ground-truth.
		& To provide standardized, reproducible, and controllable testing environments (both real and synthetic) for training, validation, and comparison. \\
		\hdashline
		
		\textbf{Integrated Protocol} (Sec.~\ref{sec_integratedprotocol}) 
		& \textbf{3-Stage Pipeline:} 1. Offline Simulation, 2. Hardware-in-the-Loop (HIL) Testing, 3. Field Testing.
		\textbf{Composite Score:} IBSA-Score (Weighted aggregation of key metrics).
		& To define a holistic testing pipeline from lab to real-world, culminating in an application-centric single score for comprehensive comparison. \\
		
		\bottomrule
	\end{tabularx}
\end{table*}
\section{Evaluation Metrics and Benchmarks}\label{sec_metrics}
After examining in the previous section the key technologies that enable {\color{black}embodied} IBSA deployment, we now turn to the construction of a holistic evaluation framework {\color{black}designed to quantify the performance of the entire IBSA architecture depicted in Fig.~\ref{fig_architecture}}, {\color{black}one that spans perception, network, and agent,} to ensure the system's behavior in real‐world deployments is quantifiable, comparable, and amenable to iteration. This section is organized around four pillars: definitions of metrics, methods of measurement, benchmark resources, and the integrated evaluation protocol.
Table~\ref{tab:evaluation_framework} summarizes this section, and its main components are detailed as follows.

\subsection{Perception Metrics}\label{sec_perceptionmetrics}
The following metrics are from~\cite{ref_DL}.
\begin{itemize}
	\item \textbf{False Positive Rate (FPR)}
	
	FPR$=$number of false alarms$/$number of negative samples.
	Alternatively, for video streams or radar point clouds, report false alarms per unit time or per unit area (e.g., per km$^2$) to characterize the system's tendency to over‐alert.
	
	\item \textbf{False Negative Rate (FNR})
	
	FNR $=$ number of missed detections $/$ number of positive samples.
	Note that Recall=1-FNR.
	
	\item \textbf{Precision and Recall}
	
	$\text{Precision}=\tfrac{\text{TP}}{\text{TP+FP}}$, $\text{Recall}=\tfrac{\text{TP}}{\text{TP+FN}}$.	
	In detection tasks, report the precision-recall curve and compute the F1 score:{\color{black} $\text{F1}=2\tfrac{\text{Precision}\times\text{Recall}}{\text{Precision}+\text{Recall}}$.}
	
	\item \textbf{Detection Error Tradeoff (DET) Curve}
	
	Plot FPR against FNR across decision thresholds to visually compare algorithmic trade‐offs under different operational priorities (e.g., higher safety vs. lower miss cost).
	
	\item \textbf{Average False‑Alarm Distance / Average Miss Duration}
	
	Incorporating target spatial-temporal distribution, define ``average false‐alarm distance" (mean distance from a false alarm to the nearest true target) or ``average miss duration" (time from target appearance to detection) to assess spatiotemporal performance in tracking and monitoring.
	
	\item \textbf{Mean Average Precision (mAP)}
	
	Compute the area under the precision-recall curve at various {\color{black}intersection over union (IoU)} thresholds. Commonly report mAP@0.5 and mAP@0.5:0.95 for multi‐class object detection.
	
	\item \textbf{Spatial Localization Error}
	
	Measure the Euclidean distance between detected object centers and ground‐truth in a global coordinate frame. Report root mean square error (RMSE) separately for near‐field and far‐field to reflect localization accuracy at different ranges.
	
	\item \textbf{2D/3D IoU}
	
	Evaluate overlap of planar bounding boxes (2D) or volumetric cuboids (3D). For 3D IoU, also account for yaw and pitch errors in voxel fitting.
	
	\item \textbf{Distance / Velocity / Angle Estimation Error}
	
	For continuously moving targets, report RMSE (and 95\% confidence intervals) for distance, speed, and attitude angle estimates, {\color{black}e.g., for multi‐step (1~s, 3~s) trajectory prediction tasks.}
	
	\item \textbf{Missing‑Rate}
	
	After sensor fusion, the fraction of dropped frames across modalities. Used to evaluate data integrity and robustness under high load or adverse conditions.
\end{itemize}
\begin{figure*}[!t]
	\centering
	\includegraphics[width=0.95\textwidth]{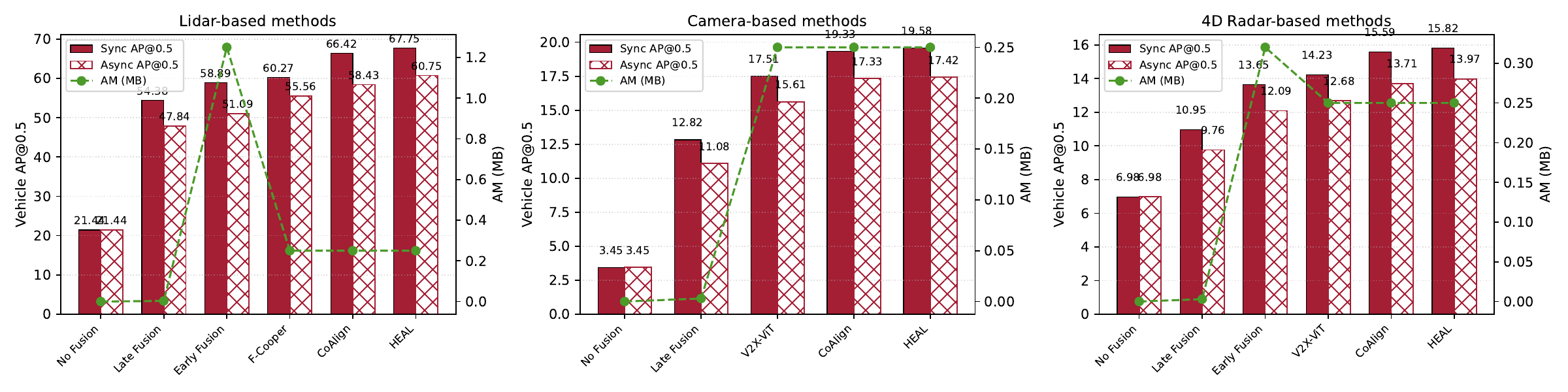}
	\caption{Performance comparison of different fusion strategies across modalities (lidar, camera, RF). Data is derived from~\cite{yang2024v2x}. The significant gap between ``No Fusion'' and cooperative methods (e.g., HEAL~\cite{lu2024heal}, CoAlign~\cite{lu2023coalign}, F-Cooper~\cite{chen2019fcooper}, V2X-ViT~\cite{xu2022v2xvit}) demonstrates the necessity of network-level collaboration for the IBSA. AP: average precision; Sync: synchronous; Aync: asynchronous; AM:average transmission cost in mega byte).}
	\label{fig:co_perception}
	
	\vspace{0.5cm} 
	
	\includegraphics[width=0.95\textwidth]{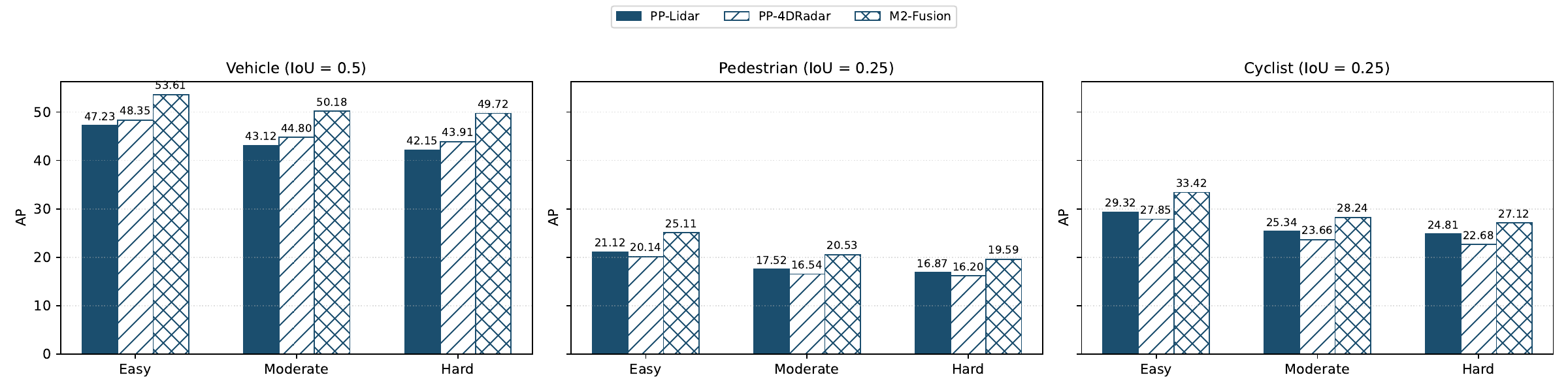}
	\caption{Ablation study on perception modalities under different difficulty levels (easy, moderate, hard). Data is derived from~\cite{yang2024v2x}. The consistent performance of radar (equivalent to wireless data of IBSA) and the superior performance of fusion models highlight the necessity of incorporating wireless sensing for robust perception. M2-Fusion comes from~\cite{yin2021centerbased}. PP: PointPillars~\cite{lang2019pointpillars}.}
	\label{fig:ablation}
	
\end{figure*}
\subsection{Network Metrics}\label{sec_networkmetrics}
{\color{black}The following metrics are from~\cite{ref_ITU}.}
\begin{itemize}
	\item \textbf{Throughput (Gbps/Hz)}
	
	Data rate per unit bandwidth, reported separately for uplink (sensor data return) and downlink (control command broadcast).
	
	\item \textbf{End‑to‑End Latency (ms)}
	
	Round‐trip delay from terminal data upload, through base‐station inference, to issuance of control commands. Emphasize the 99.9th percentile to capture tail latency.
	
	\item \textbf{Energy per Inference (J)}
	
	Energy consumed for each fused perception‐inference task. Break down total into GPU, RF front end, digital signal processor (DSP), etc.
	
	\item {\color{black}\textbf{Interference Level}}
	
	Quantify the impact of {\color{black}embodied} IBSA's RF perception or jamming operations on legacy communications, e.g., via signal-to-interference-plus-noise ratio (SINR) degradation or block error rate (BLER) increase in neighboring cells.
\end{itemize}

\subsection{Agent‑Level Metrics}\label{sec_agentmetrics}

\begin{itemize}
	\item {\color{black}\textbf{Task Completion Rate (TCR)~\cite{ref_TCR}}}
	
	Proportion of closed‐loop {\color{black}embodied} IBSA processes, from object detection and beamforming/jamming decisions to coordinated control, that successfully meet safety and real‐time constraints.
	
	\item \textbf{Robustness Drop}
	
	Performance degradation (in mAP or TCR) under adverse conditions (e.g., fog, occlusion, channel fading) relative to baseline.
\end{itemize}

\subsection{Benchmark Resources}\label{sec_benchmark}
To ensure repeatability and comparability, the {\color{black}embodied} IBSA evaluation framework leverages both public multimodal datasets and an end‑to‑end digital‐twin simulation platform. Public datasets provide annotated real‐world samples (vision, radar, wireless channels)~\cite{yang2024v2x}, while simulation tools (e.g., CARLA, Sionna) generate controllable, customizable synchronized sensor data and ground‐truth trajectories~\cite{ref_DeepVerse}. Together, they comprehensively support perception, network, and agent metrics.

Notably, the V2X-Radar dataset~\cite{yang2024v2x} provides critical benchmarks for cooperative perception, offering quantitative evidence for the design principles of the IBSA architecture. As visualized in Fig.~\ref{fig:co_perception}, comparisons between ``No Fusion'' and various cooperative strategies demonstrate that single-node perception is insufficient for complex environments, whereas multi-source fusion yields significant performance gains. This validates the necessity of the IBSA's collaborative network structure. Furthermore, regarding the sensing modality, Fig.~\ref{fig:ablation} illustrates that while optical sensors are susceptible to degradation, the inclusion of wireless sensing data is essential for maintaining robustness under varying difficulty levels. This supports the IBSA's core premise of utilizing wireless signals for resilient environmental perception.

\subsection{Integrated Evaluation Protocol}\label{sec_integratedprotocol}
To jointly optimize {\color{black}individual agent and overall network metrics}, we propose a four‑stage testing pipeline with clearly defined terminology and procedures at each step:
\begin{itemize}
	\item \textbf{Offline Simulation}
	
	Generate ``standard conditions'' (normal traffic, good channels) and ``extreme conditions" (harsh weather, high interference) in the digital twin.	
	Perception, communication/network, and compuation metrics for pretrained and fine‐tuned models to establish {\color{black}baselines~\cite{ref_Unreal}.}
	
	\item \textbf{Hardware-in-the-Loop Testing (HIL)}
	
	Deploy models to a prototype BS integrating software-defined radio (SDR) and GPU.	
	In a closed‐lab setup, emulate real‐world links, record inference latency, GPU/RF power curves, and network throughput/interference metrics in {\color{black}real time~\cite{ref_DeepVerse}.}
	
	\item \textbf{Field Testing}
	
	Co‐deploy {\color{black}embodied} IBSA and a traditional evolved node B (eNB) (or other baseline) at the same test site.	
	Compare TCR, {\color{black}99.9th percentile latency}, $\Delta$SINR/$\Delta$BLER under identical traffic and channel conditions to quantify real‐world gains.
	
	\item \textbf{Composite {\color{black}Embodied} IBSA‑Score}
	
	{\color{black}To yield a single comprehensive performance indicator, we propose the composite IBSA-Score.} We define this score by aggregating $N$ core metrics $m_i$ (detection accuracy, Recall, mAP, 1$/$center-error, normalized throughput, 1$/$latency, 1$/$energy, TCR) with predefined weights:
	$\text{IBSA-Score} = \sum_{i=1}^{N} w_i m_i, \quad \sum_{i=1}^{N} w_i = 1.$
	Weights can be tuned per application scenario (e.g., 
	\emph{safety} $>$ \emph{latency} $>$ \emph{throughput}) 
	to yield a single comprehensive performance indicator.
	This tuning is intentionally application-centric, reflecting operational priorities, much like setting critical hyperparameters in a machine learning model based on the end goal. While a deep exploration of specific weighting methodologies is outside the scope of this review, formal methods such as the analytic hierarchy process could be employed to systematically derive these weights based on operator-defined pairwise comparisons of the metrics' importance.
\end{itemize}

\section{Challenges and Future Directions}\label{sec_open}
Building on the metrics and evaluation protocol introduced above, we now pivot from what to measure to what remains hard, outlining open problems and concrete directions for the {\color{black}embodied} IBSA research community.

\subsection{Advancing the Cognitive Core: Robustness, Adaptation, and Action}
\subsubsection{Robust Multimodal Perception Algorithms}  
Cross-modal alignment under occlusion, glare or spectrum scarcity demands architectures that speak \lq\lq RF\,+\,vision\,+\,point-cloud\rq\rq.  Transformers such as {\color{black}MVX-ViT in~\cite{ref_MVX}} encode radio tokens alongside image patches and use spatio-temporal attention to fuse beam-space priors with dense semantics, reducing missed detections and improving beam prediction accuracy in V2X scenes.  Research frontiers now include causal fusion, calibrated uncertainty and on-device distillation so that perception meets millisecond edge budgets.

\subsubsection{Few-Shot Adaptation \& Lifelong Learning}  
Real roads and air corridors evolve faster than any static dataset.  Parameter-efficient fine-tuning (LoRA, adapters) lets each BS personalise {\color{black}the core LAM} with only kilobytes of updates, while federated aggregation guards privacy under {\color{black}non-independent and non-identically} distributed traffic.
Open problems include catastrophic forgetting, poisoned updates, and balancing exploration-exploitation in continually shifting wireless channels.

\subsubsection{LAM-Driven Executable Policies via Reinforcement Learning}
LAMs can leverage reinforcement learning to translate their rich world models into sequential decision-making capabilities. This allows an agent to convert perception outputs into safe, executable actions, such as beamforming, jamming, or path-planning, that adhere to strict QoS and legal constraints. Key challenges in operationalizing this approach lie in the effective use of offline RL with telecom digital twins, the design of reward functions that properly balance throughput and safety, and the development of formally verified policy updates for mission-critical links~\cite{Yue2024}.

\subsection{System Realization: Optimization, Energy, and Simulation}
\subsubsection{Joint Perception-Communication-Computation Optimisation}\label{subsection5}  
Perception accuracy, spectrum use and GPU cycles are tightly coupled.  Cross-layer Lagrangian solvers and graph-RL schedulers have been proposed to minimise {\color{black}99.9th percentile} latency while maximising mAP and throughput, yet real-time feedback between the {\color{black}physical layer} and application layers is still missing.  Priority pre-emption for bursty events (e.g., rogue drones) and carbon-aware scheduling are fertile grounds for exploration.

\subsubsection{Energy Modelling and Resource Efficiency}  
Fine-grained power attribution for RF front-ends and accelerators is largely absent.  Accurate joule-per-inference models would unlock dynamic voltage/frequency scaling, smart sleeping of antenna panels and carbon-aware orchestration.  Paired with the optimisation in Subsection~\ref{subsection5}, this enables longer UAV patrol endurance and greener 6G infrastructure.

\subsubsection{Digital Twins and Sim-to-Real Transfer}  
Photo-realistic engines (CARLA~\cite{ref_carla}, AirSim~\cite{ref_AirSim}) plus RF ray-tracers (Sionna~\cite{ref_sionna}, Wireless InSite~\cite{ref_WirelessInSite}) already support closed-loop beam-selection studies, but residual gaps persist in urban-canyon multipath and extreme weather.  Domain randomisation, generative replay and hybrid hardware-in-the-loop prototypes are required to guarantee that policies generalise from \lq\lq silicon city\rq\rq{} to real asphalt.

\subsection{Standardization and Evaluation Ecosystem}
\subsubsection{Unified Benchmarks \& Open Datasets}  
A lasting obstacle for embodied integrated perception, communication, and computating and {\color{black}embodied} IBSA research is the absence of universally accepted datasets and KPI suites.  Public corpora such as {\color{black}DeepSense-6G in~\cite{ref_DeepSense}} already combine mmWave CSI, lidar and camera data, showing how multi-modal ground truth can be collected at scale.  A single composite metric, {\color{black}e.g.,} the {\color{black}embodied} IBSA-Score that blends perception, network and agent KPIs, must be elevated from lab script to community leaderboard.  Future work should extend benchmarks to low-altitude UAV safety and long-tail driving, release time-synchronised raw RF captures, and publish hardware-in-the-loop test scripts so that academic code and 3rd generation partnership Project (3GPP) conformance reports become mutually reproducible.

A critical goal for these future benchmarks will be to provide quantitative comparisons between the proposed IBSA architecture and existing 5G-advanced ISAC BSs. Although IBSA may incur higher computational overhead due to its LAM core, its expected performance gains are significant. For instance, higher mAP is expected through robust multimodal fusion (e.g., RF plus vision) that overcomes the limitations of RF-only sensing in adverse conditions~\cite{yang2024v2x}. Furthermore, lower perceived latency is anticipated, enabled by the tight coupling of the perception-cognition-execution loop at the edge, which allows for millisecond-level local actions without core-network round trips. Finally, system-level energy efficiency should be realized via joint perception-communication-computation co-optimization and resource slicing, aiming to minimize the energy-per-task rather than just raw hardware power. Establishing public testbeds to validate these hypotheses, using the IBSA-Score as a holistic metric, remains a crucial next step.

\subsubsection{Standardisation and Industry Momentum}  
Scaling from prototypes to city-wide roll-outs demands a single, interoperable playbook.
	Both 3GPP and ETSI are making critical progress. 3GPP's study on ISAC (TR 22.837~\cite{ref_3GPP_ISAC}) is defining shared channel models and perception-first KPIs across bands. Concurrently, ETSI's industry specification group (ISG) on ISAC and its work on AI-native systems are establishing foundational use cases and architecture principles~\cite{ETSIISAC}.
	
	From a compatibility perspective, the IBSA architecture aligns with these efforts but introduces specific challenges that require further standardization. For instance, IBSA's reliance on multimodal fusion (RF, visual, lidar), as discussed in this paper, requires data formats and synchronization protocols that go beyond the RF-centric perception data currently prioritized in TR 22.837. Similarly, the ``cognition" layer of IBSA necessitates standardized interface protocols for LAM lifecycle management (like federated updates and MoE routing ), which expands on 3GPP's existing AI for new radio work-item~\cite{ref_3GPP_AI}.
	
	Therefore, a key future task is to bridge this gap. Aligning academic datasets and the proposed composite embodied IBSA-Score  with these emerging 3GPP and ETSI drafts is essential. This will help evolve IBSA from a research concept into a standardized, interoperable component, accelerating multi-vendor interoperability and mass adoption.

\subsection{Trustworthy AI: Explainability, Governance, and Ethics}
\subsubsection{Explainability and Trustworthiness}
Explainability is a fundamental requirement, as safety regulators demand a verifiable causal link for every system action~\cite{ldrdd2025}. Techniques like self-explainable neural layers and attention heat-maps can clarify how the model aligns radio-frequency and visual data. Furthermore, counter-factual reasoning and signed execution logs are essential for verifying decisions and exploring alternative outcomes. These mechanisms build trust with operators and the public while forming a critical defense against adversarial attacks, such as manipulated RF signals or spoofed visual markers.

\subsubsection{Ethics, Governance, and Fairness}  
Massive multi-sensor surveillance raises concerns over privacy, equitable spectrum sharing and low-altitude air-traffic control.  Embedding differential-privacy guards, audit trails and human-in-the-loop overrides into network fabric must become a design primitive.  Only by coupling technical excellence with transparent governance can 6G embodied agents earn societal acceptance.

\section{Conclusion}\label{sec_conclusion}
This paper explored how {\color{black}LAMs} can enable {\color{black}embodied intelligence} in wireless infrastructure by transforming BSs into {\color{black}embodied} IBSAs. We presented a three-layer perception-cognition-execution architecture that integrates multimodal perception with policy generation and controllable actions, implemented via cloud-edge-end collaboration and parameter-efficient adaptation. Two application cases, {\color{black}cooperative vehicle-road perception for autonomous driving and low-altitude safety against unauthorized UAVs,} illustrated how {\color{black}embodied} IBSAs unlock complementary strengths across radio, vision, and computation to improve both safety and connectivity.

To support rigorous progress, we organized enabling technologies (cognitive core, system architecture, and security as well as trustworthiness) and proposed an evaluation framework that unifies {\color{black}individual agent and overall network metrics}. A composite {\color{black}embodied} IBSA-Score was outlined to facilitate apples-to-apples comparisons and to connect research prototypes with engineering requirements.

Looking ahead, the most pressing needs are: (i) open, time-synchronized multimodal datasets that include raw RF captures and hardware-in-the-loop artifacts; (ii) robust, uncertainty-aware fusion and continual learning that survive occlusion, domain shift, and adversarial conditions; (iii) formally verifiable policy execution that respects QoS, safety, and legal constraints; (iv) joint perception-communication-computation scheduling under tight energy budgets; and (v) standardization of interfaces, metrics, and model lifecycle management, ensuring IBSA's multimodal data formats and API protocols align with emerging frameworks from bodies like 3GPP (e.g., TR 22.837) and ETSI ISG-ISAC, coupled with transparent governance. Addressing these challenges will accelerate reproducible, trustworthy deployment of LAM-enabled embodied IBSAs and help establish integrated perception, communication, and computation-native intelligence as a practical cornerstone of 6G networks.
\label{key}


\bibliographystyle{cjereport}
\bibliography{refs}

\begin{thebibliography}{100}
\newcommand{\enquote}[1]{``#1''}
\providecommand{\url}[1]{\texttt{#1}}
\providecommand{\urlprefix}{ }
\expandafter\ifx\csname urlstyle\endcsname\relax
  \providecommand{\doi}[1]{doi:\discretionary{}{}{}#1}\else
  \providecommand{\doi}{doi:\discretionary{}{}{}\begingroup
  \urlstyle{rm}\Url}\fi
\providecommand{\eprint}[2][]{\url{#2}}

\bibitem{ref_NativeAI_3}
X.~You, Y.~Huang, C.~Zhang, \emph{et~al.}, \enquote{When {AI} meets sustainable
  {6G}}, \emph{Sci. China Inf. Sci.}, vol.68, artilce no.110301,
  \doi{10.1007/s11432-024-4257-6}, 2025.

\bibitem{ref_NativeAI_2}
W.~Saad, O.~Hashash, C.~K. Thomas, C.~Chaccour, \emph{et~al.},
  \enquote{Artificial general intelligence ({AGI})-native wireless systems: A
  journey beyond {6G}}, \emph{Proc. IEEE}, in press, pp.1--39,
  \doi{10.1109/JPROC.2025.3526887}, early Access, 2025.

\bibitem{ref_Huang2024}
Y.~Huang, \enquote{Challenges and opportunities of sub-6 {GHz} integrated
  sensing and communications for {5G-Advanced} and beyond}, \emph{Chin. J.
  Electron.}, vol.33, no.2, pp.323--325, \doi{10.23919/cje.2023.00.251}, 2024.

\bibitem{ref_ISAC}
F.~Liu, Y.~Cui, C.~Masouros, J.~Xu, \emph{et~al.}, \enquote{Integrated sensing
  and communications: Toward dual-functional wireless networks for {6G} and
  beyond}, \emph{IEEE J. Sel. Areas Commun.}, vol.40, no.6, pp.1728--1767,
  \doi{10.1109/JSAC.2022.3156632}, 2022.

\bibitem{ref_xf_JSAC}
X.~Fang, C.~Lei, W.~Feng, Y.~Chen, \emph{et~al.},
  \enquote{Sensing-communication-computing-control closed-loop optimization for
  {6G} digital twin-empowered robotic systems}, \emph{IEEE J. Sel. Areas
  Commun.}, vol.43, no.10, pp.3330--3346, \doi{10.1109/JSAC.2024.3465133},
  2025.

\bibitem{ref_cl_Network}
C.~Lei, X.~Fang, W.~Feng, Y.~Chen, \emph{et~al.}, \enquote{Satellite-{UAV}
  networks for {6G} control: A sensing-communication-computing-control closed
  loop perspective}, \emph{IEEE Netw.}, vol.39, no.4, pp.62--69,
  \doi{10.1109/MNET.001.2400192}, 2025.

\bibitem{ref_cl_JSAC}
C.~Lei, W.~Feng, P.~Wei, Y.~Chen, \emph{et~al.}, \enquote{Edge information hub:
  Orchestrating satellites, {UAVs}, {MEC}, sensing and communications for {6G}
  closed-loop controls}, \emph{IEEE J. Sel. Areas Commun.}, vol.43, no.1,
  pp.5--20, \doi{10.1109/JSAC.2024.3461414}, 2025.

\bibitem{luhme2025proc}
H.~L. Cardoso, R.~Sousa-Silva, M.~Koponen, and A.~Pareja-Lora (Eds.),
  \emph{Proceedings of the 2nd {LUHME} Workshop}, {LUHME}, Bologna, Italy,
  2025, ISBN 978-989-9193-73-4, \doi{10.21747/978-989-9193-73-4/lan2},
  \urlprefix\url{https://aclanthology.org/2025.luhme-1.pdf}.

\bibitem{zhou2025perception}
C.~Zhou, M.~Wang, Y.~Ma, C.~Wu, \emph{et~al.}, \emph{\enquote{From perception
  to cognition: A survey of vision-language interactive reasoning in multimodal
  large language models}, arXiv e-Print}, arXiv:2509.25373, 2025.

\bibitem{IoTJMFengZhiyong}
A.~Liu, W.~Jiang, S.~Huang, and Z.~Feng, \enquote{Multi-modal integrated
  sensing and communication in internet of things with large language models},
  \emph{IEEE Internet Things J. Mag.}, vol.8, no.5, pp.103--111,
  \doi{10.1109/MIOT.2025.3575888}, 2025.

\bibitem{ref_LAM_2}
H.~Zou, Q.~Zhao, Y.~Tian, L.~Bariah, \emph{et~al.}, \enquote{{TelecomGPT}: A
  framework to build telecom-specific large language models}, \emph{IEEE Trans.
  Mach. Learn. Commun. Netw.}, vol.3, pp.948--975,
  \doi{10.1109/TMLCN.2025.3593184}, 2025.

\bibitem{ref_LAM_3}
F.~Zhu, X.~Wang, X.~Li, M.~Zhang, \emph{et~al.}, \emph{\enquote{Wireless large
  {AI} model: Shaping the {AI}-native future of {6G} and beyond}, arXiv
  e-Print}, arXiv:2504.14653, 2025.

\bibitem{ref_LAM_1}
Z.~Wang, Y.~Shi, Y.~Zhou, J.~Zhu, and K.~B. Letaief, \enquote{Edge large {AI}
  models: Revolutionizing {6G} networks}, \emph{IEEE Commun. Mag.}, vol.63,
  no.10, pp.36--42, \doi{10.1109/MCOM.001.2400761}, 2025.

\bibitem{ref_DeepSense}
A.~Alkhateeb, G.~Charan, T.~Osman, A.~Hredzak, \emph{et~al.},
  \enquote{{DeepSense 6G}: A large-scale real-world multi-modal sensing and
  communication dataset}, \emph{IEEE Commun. Mag.}, vol.61, no.9, pp.122--128,
  \doi{10.1109/MCOM.006.2200730}, 2023.

\bibitem{ref_DeepVerse}
W.~I. Lab, \enquote{Deepverse {6G}: A digital twin dataset for wireless
  communication and sensing}, 2025, \urlprefix\url{https://deepverse6g.net/}.

\bibitem{ref_Unreal}
K.~Huang, S.~Mu, J.~Jiang, Y.~Gao, and S.~Xu, \emph{\enquote{Unreal is all you
  need: Multimodal {ISAC} data simulation with only one engine}, arXiv
  e-Print}, arXiv:2507.08716, 2025.

\bibitem{ref_MVX}
G.~Gharsallah and G.~Kaddoum, \enquote{{MVX-ViT}: Multimodal collaborative
  perception for {6G V2X} network management decisions using vision
  transformer}, \emph{IEEE Open J. Commun. Soc.}, vol.5, pp.5619--5634,
  \doi{10.1109/OJCOMS.2024.3452591}, 2024.

\bibitem{ref_M3SC}
X.~Cheng, Z.~Huang, L.~Bai, H.~Zhang, \emph{et~al.}, \enquote{{M3SC}: A generic
  dataset for mixed multi-modal {(MMM)} sensing and communication integration},
  \emph{China Commun.}, vol.20, no.11, pp.13--29,
  \doi{10.23919/JCC.fa.2023-0268.202311}, 2023.

\bibitem{ref_SoM}
X.~Cheng, Z.~Huang, L.~Yu, Yong~and Bai, M.~Sun, \emph{et~al.},
  \enquote{{SynthSoM}: A synthetic intelligent multi-modal
  sensing-communication dataset for synesthesia of machines {(SoM)}},
  \emph{Sci. Data}, vol.12, artilce no.819, \doi{10.1038/s41597-025-05065-x},
  2025.

\bibitem{ref_TVT_dataset}
B.~Salehi, G.~Reus-Muns, D.~Roy, Z.~Wang, \emph{et~al.}, \enquote{Deep learning
  on multimodal sensor data at the wireless edge for vehicular network},
  \emph{IEEE Trans. Veh. Technol.}, vol.71, no.7, pp.7639--7655,
  \doi{10.1109/TVT.2022.3170733}, 2022.

\bibitem{ref_Infocom_Dataset}
S.~Pradhan, D.~Roy, B.~Salehi, and K.~Chowdhury, \enquote{{COPILOT}:
  Cooperative perception using lidar for handoffs between road side units}, in
  \emph{Proc. IEEE Conf. Comput. Commun. (INFOCOM)}, pp.1301--1310,
  \doi{10.1109/INFOCOM52122.2024.10621174}, 2024.

\bibitem{ref_TMC_Dataset}
D.~Muruganandham, S.~Pradhan, J.~Gu, T.~Braun, \emph{et~al.}, \enquote{{SMART}:
  {Sim2Real} meta-learning-based training for mmwave beam selection in {V2X}
  networks}, \emph{IEEE Trans. Mob. Comput.}, vol.24, no.10, pp.1--16,
  \doi{10.1109/TMC.2025.3576203}, 2025.

\bibitem{ref_WC_Dataset}
S.~Kim, J.~Moon, J.~Kim, Y.~Ahn, \emph{et~al.}, \enquote{Role of sensing and
  computer vision in {6G} wireless communications}, \emph{IEEE Wirel. Commun.},
  vol.31, no.5, pp.264--271, \doi{10.1109/MWC.016.2300526}, 2024.

\bibitem{ref_HADAR}
F.~Bao, X.~Wang, S.~H. Sureshbabu, G.~Sreekumar, \emph{et~al.},
  \enquote{Heat-assisted detection and ranging}, \emph{Nature}, vol.619,
  pp.743--747, \doi{10.1038/s41586-023-06174-6}, 2023.

\bibitem{yang2024v2x}
L.~Yang, X.~Zhang, J.~Li, C.~Wang, \emph{et~al.}, \enquote{{V2X-Radar}: A
  multi-modal dataset with {4D} radar for cooperative perception}, \emph{Adv.
  Neural Inf. Process. Syst. (NeurIPS)}, in press, 2025.

\bibitem{mao2025multimodal}
T.~Mao, L.~Liang, J.~Yang, H.~Ye, \emph{et~al.}, \enquote{Multimodal-wireless:
  A large-scale dataset for sensing and communication}, in press,
  no.2511.03220, 2025.

\bibitem{ref_24_arXiv_DeepSense_Predict}
G.~Charan and A.~Alkhateeb, \emph{\enquote{Sensing-aided {6G} drone
  communications: Real-world datasets and demonstration}, arXiv e-Print},
  arXiv:2412.04734, 2024.

\bibitem{ref_25_arXiv_carla_sionna}
Y.~M. Park, Y.~K. Tun, W.~Saad, and C.~S. Hong,
  \emph{\enquote{Resource-efficient beam prediction in {mmWave} communications
  with multimodal realistic simulation framework}, arXiv e-Print},
  arXiv:2504.05187, 2025.

\bibitem{ref_VTCFall_carla}
L.~Cazzella, F.~Linsalata, M.~Magarini, M.~Matteucci, and U.~Spagnolini,
  \enquote{A multi-modal simulation framework to enable digital twin-based
  {V2X} communications in dynamic environments}, in \emph{2024 IEEE 100th
  Vehicular Technology Conference (VTC2024-Fall)}, pp.1--6,
  \doi{10.1109/VTC2024-Fall63153.2024.10757947}, 2024.

\bibitem{ref_TMLCN}
G.~Gharsallah and G.~Kaddoum, \enquote{Multimodal collaborative perception for
  dynamic channel prediction in {6G V2X} networks}, \emph{IEEE Trans. Mach.
  Learn. Commun. Netw.}, vol.3, pp.725--743, \doi{10.1109/TMLCN.2025.3578577},
  2025.

\bibitem{ref_TN}
B.~Salehi, U.~Demir, D.~Roy, S.~Pradhan, \emph{et~al.}, \enquote{Multiverse at
  the edge: Interacting real world and digital twins for wireless beamforming},
  \emph{IEEE/ACM Trans. on Netw.}, vol.32, no.4, pp.3092--3110,
  \doi{10.1109/TNET.2024.3377114}, 2024.

\bibitem{ref_Embodied_1}
H.~Liu, D.~Guo, and A.~Cangelosi, \enquote{Embodied intelligence: A synergy of
  morphology, action, perception and learning}, \emph{ACM Comput. Surv.},
  vol.57, no.7, ISSN 0360-0300, 2025.

\bibitem{ref_Embodied_2}
X.~Wang, F.~Zhu, Z.~Yang, C.~Huang, \emph{et~al.}, \emph{\enquote{Bridging
  physical and digital worlds: Embodied large {AI} for future wireless
  systems}, arXiv e-Print}, arXiv:2506.24009, 2025.

\bibitem{ref_Embodied_3}
S.~Ruan, R.~Wang, X.~Shen, H.~Liu, \emph{et~al.}, \emph{\enquote{A survey of
  multi-sensor fusion perception for embodied {AI}: Background, methods,
  challenges and prospects}, arXiv e-Print}, arXiv:2506.19769, 2025.

\bibitem{ref_UE}
{Epic Games}, \enquote{{Unreal Engine} (version 5.6)}, [Software]
  \url{https://www.unrealengine.com/}, 2025.

\bibitem{ref_carla}
A.~Dosovitskiy, G.~Ros, F.~Codevilla, A.~Lopez, and V.~Koltun,
  \enquote{{CARLA}: {An} open urban driving simulator}, in \emph{Proc. Conf.
  Robot Learn. (CoRL)}, pp.1--16, 2017.

\bibitem{ref_blender}
{Blender Development Team}, \enquote{{Blender} (version 4.5)}, [Computer
  software] \url{https://www.blender.org/}, 2022.

\bibitem{ref_blensor}
{Blensor Project}, \enquote{{BlenSor}: Blender sensor simulation toolbox},
  \url{https://www.blensor.org/}.

\bibitem{ref_AirSim}
S.~Shah, D.~Dey, C.~Lovett, and A.~Kapoor, \enquote{{AirSim}: High-fidelity
  visual and physical simulation for autonomous vehicles}, in \emph{Field and
  Service Robotics}, 1705.05065,
  \urlprefix\url{https://arxiv.org/abs/1705.05065}, 2017.

\bibitem{ref_OpenGL}
{Khronos Group}, \enquote{{OpenGL} 4.6 specification},
  \urlprefix\url{https://www.khronos.org/registry/OpenGL/specs/gl/glspec46.core.pdf}.

\bibitem{ref_WirelessInSite}
{Remcom}, \enquote{{Wireless InSite}\textregistered{} 3d wireless prediction
  software (version 4.0)}, [Software]
  \url{https://www.remcom.com/wireless-insite-propagation-software}, 2025.

\bibitem{ref_sionna}
J.~Hoydis, S.~Cammerer, F.~{Ait Aoudia}, M.~Nimier-David, \emph{et~al.},
  \enquote{Sionna}, [Software] \url{https://nvlabs.github.io/sionna/}, 2022.

\bibitem{ref_WaveFarer}
{Remcom}, \enquote{{WaveFarer} radar simulation software (version 2.1.0.6)},
  [Software] \url{https://www.remcom.com/wavefarer-automotive-radar-software},
  2025.

\bibitem{ref_24_Bosch_radar}
K.~Armanious, M.~Quach, M.~Ulrich, T.~Winterling, \emph{et~al.},
  \emph{\enquote{Bosch street dataset: A multi-modal dataset with imaging radar
  for automated driving}, arXiv e-Print}, arXiv:2407.12803, 2024.

\bibitem{ref_24_TIV_radar_camera}
S.~Yao, R.~Guan, X.~Huang, Z.~Li, \emph{et~al.}, \enquote{Radar-camera fusion
  for object detection and semantic segmentation in autonomous driving: A
  comprehensive review}, \emph{IEEE Trans. Intell. Veh.}, vol.9, no.1,
  pp.2094--2128, \doi{10.1109/TIV.2023.3307157}, 2024.

\bibitem{ref_24_TITS_lidar_camera}
X.~Wang, K.~Li, and A.~Chehri, \enquote{Multi-sensor fusion technology for {3D}
  object detection in autonomous driving: A review}, \emph{IEEE Trans. Intell.
  Transp. Syst.}, vol.25, no.2, pp.1148--1165, \doi{10.1109/TITS.2023.3317372},
  2024.

\bibitem{ref_23_ICRA_lidar_camera}
Z.~Liu, H.~Tang, A.~Amini, X.~Yang, \emph{et~al.}, \enquote{{BEVFusion}:
  Multi-task multi-sensor fusion with unified bird's-eye view representation},
  in \emph{Proc. IEEE Int. Conf. Robot. Autom. (ICRA)}, pp.2774--2781,
  \doi{10.1109/ICRA48891.2023.10160968}, 2023.

\bibitem{ref_25_TITS_radar_camera}
H.~Liu, J.~Liu, G.~Jiang, and X.~Jin, \enquote{{MSSF}: A {4D} radar and camera
  fusion framework with multi-stage sampling for {3D} object detection in
  autonomous driving}, \emph{IEEE IEEE Trans. Intell. Transp. Syst.}, vol.26,
  no.6, pp.8641--8656, \doi{10.1109/TITS.2025.3554313}, 2025.

\bibitem{ref_24_IROS_radar_camera}
N.~Baumann, M.~Baumgartner, E.~Ghignone, J.~Kühne, \emph{et~al.},
  \enquote{{CR3DT}: Camera-radar fusion for {3D} detection and tracking}, in
  \emph{Proc. IEEE/RSJ Int. Conf. Intell. Robot. Syst. (IROS)}, pp.4926--4933,
  \doi{10.1109/IROS58592.2024.10801848}, 2024.

\bibitem{ref_gz_1}
Y.~Mei, Z.~Gao, Y.~Wu, W.~Chen, \emph{et~al.}, \enquote{Compressive
  sensing-based joint activity and data detection for grant-free massive {IoT}
  access}, \emph{IEEE Trans. Wireless Commun.}, vol.21, no.3, pp.1851--1869,
  \doi{10.1109/TWC.2021.3107576}, 2022.

\bibitem{ref_gz_2}
X.~Zhou, K.~Ying, Z.~Gao, Y.~Wu, \emph{et~al.}, \enquote{Active terminal
  identification, channel estimation, and signal detection for grant-free
  {NOMA-OTFS} in {LEO} satellite internet-of-things}, \emph{IEEE Trans.
  Wireless Commun.}, vol.22, no.4, pp.2847--2866,
  \doi{10.1109/TWC.2022.3214862}, 2023.

\bibitem{ref_gz_3}
K.~Ying, Z.~Gao, S.~Chen, M.~Zhou, \emph{et~al.}, \enquote{Quasi-synchronous
  random access for massive {MIMO}-based {LEO} satellite constellations},
  \emph{IEEE J. Sel. Areas Commun.}, vol.41, no.6, pp.1702--1722,
  \doi{10.1109/JSAC.2023.3273699}, 2023.

\bibitem{ref_LZR_JSAC}
Z.~Li, Z.~Gao, B.~Ning, and Z.~Wang, \enquote{Radiation pattern reconfigurable
  {FAS}-empowered interference-resilient uav communication}, \emph{IEEE J. Sel.
  Areas Commun.}, in press, pp.1--1, \doi{10.1109/JSAC.2025.3617928}, 2025.

\bibitem{ref_lzr_JSTSP}
Z.~Li, Z.~Gao, and T.~Li, \enquote{Sensing user's channel and location with
  terahertz extra-large reconfigurable intelligent surface under hybrid-field
  beam squint effect}, \emph{IEEE J. Sel. Top. Signal Process.}, vol.17, no.4,
  pp.893--911, \doi{10.1109/JSTSP.2023.3278942}, 2023.

\bibitem{ref_lzr_IoTJ}
Z.~Li, Z.~Gao, K.~Wang, Y.~Mei, \emph{et~al.}, \enquote{Unauthorized {UAV}
  countermeasure for low-altitude economy: Joint communications and jamming
  based on {MIMO} cellular systems}, \emph{IEEE Internet Things J.}, vol.12,
  no.6, pp.6659--6672, \doi{10.1109/JIOT.2024.3491796}, 2025.

\bibitem{ref_lzr_Network}
Z.~Li, S.~Tan, Z.~Gao, Y.~Tao, \emph{et~al.}, \enquote{Chirp delay-doppler
  domain modulation: A new paradigm of integrated sensing and communication for
  autonomous vehicles}, \emph{IEEE Netw.}, vol.39, no.6, pp.119--127,
  \doi{10.1109/MNET.2025.3573832}, 2025.

\bibitem{ref_zxy_JSAC}
Z.~Gao, X.~Zhou, B.~Ning, Y.~Su, \emph{et~al.}, \enquote{Integrated location
  sensing and communication for ultra-massive {MIMO} with hybrid-field
  beam-squint effect}, \emph{IEEE J. Sel. Areas Commun.}, vol.43, no.4,
  pp.1387--1404, \doi{10.1109/JSAC.2025.3531551}, 2025.

\bibitem{jiang2025comprehensive}
F.~Jiang, C.~Pan, L.~Dong, K.~Wang, \emph{et~al.}, \emph{\enquote{A
  comprehensive survey of large {AI} models for future communications:
  Foundations, applications and challenges}, arXiv e-Print}, arXiv:2505.03556,
  2025.

\bibitem{abel2025large}
M.~Abel, I.~Ahmad, C.~A. Casado, R.~Berner, \emph{et~al.}, {``\emph{Large
  Language Models in the {6G}-Enabled Computing Continuum: A White Paper}''},
  \emph{Doctor's thesis}, University of Oulu, 2025.

\bibitem{jiang2025large}
F.~Jiang, C.~Pan, L.~Dong, K.~Wang, \emph{et~al.}, \emph{\enquote{From large
  {AI} models to agentic {AI}: A tutorial on future intelligent
  communications}, arXiv e-Print}, arXiv:2505.22311, 2025.

\bibitem{cui2025integrated}
Y.~Cui, J.~Nie, F.~Liu, W.~Yuan, \emph{et~al.}, \emph{\enquote{Integrated
  sensing and communication: Towards multifunctional perceptive network}, arXiv
  e-Print}, arXiv:2510.14358, 2025.

\bibitem{chen2025wirelessNativeBigAI}
Z.~Chen, Z.~Zhang, C.~Liu, and Z.~Xing, \enquote{Towards wireless native big
  {AI} model: The mission and approach differ from large language model},
  \emph{Sci. China Inf. Sci.}, vol.68, no.7, artilce no.170303,
  \doi{10.1007/s11432-024-4464-8}, 2025.

\bibitem{Qin2024}
G.~Qin, X.~Meng, T.~Wen, and B.~Cai, \enquote{Virtual coupling trains based on
  multi-agent system under communication delay}, \emph{Chin. J. Electron.},
  vol.33, no.6, pp.1545--1554, \doi{10.23919/cje.2022.00.253}, 2024.

\bibitem{ref_space_1}
Z.~Gao, D.~Mi, C.~Jiang, S.~Chatzinotas, \emph{et~al.}, \enquote{Emerging space
  communication and network technologies for sixth-generation ubiquitous
  connectivity}, \emph{Space: Science \& Technology}, vol.4, artilce no.0239,
  \doi{10.34133/space.0239},
  \urlprefix\url{https://spj.science.org/doi/abs/10.34133/space.0239}, 2024.

\bibitem{ref_space_2}
X.~Liu, Z.~Gao, Z.~Wan, Z.~Wu, \emph{et~al.}, \enquote{Toward near-space
  communication network in the {6G} and beyond era}, \emph{Space: Science \&
  Technology}, vol.5, artilce no.0337, \doi{10.34133/space.0337},
  \urlprefix\url{https://spj.science.org/doi/abs/10.34133/space.0337}, 2025.

\bibitem{ref_space_3}
H.~Liu, T.~Qin, null, T.~Mao, \emph{et~al.}, \enquote{Near-space
  communications: The last piece of {6G} space–air–ground–sea integrated
  network puzzle}, \emph{Space: Science \& Technology}, vol.4, artilce no.0176,
  \doi{10.34133/space.0176},
  \urlprefix\url{https://spj.science.org/doi/abs/10.34133/space.0176}, 2024.

\bibitem{ref_Zhang2025}
Z.~Zhang, R.~He, M.~Yang, X.~Zhang, \emph{et~al.}, \enquote{Non-stationarity
  characteristics in dynamic vehicular {ISAC} channels at 28 {GHz}},
  \emph{Chin. J. Electron.}, vol.34, no.1, pp.73--81,
  \doi{10.23919/cje.2024.00.003}, 2025.

\bibitem{ref_MultiX}
{MultiX Consortium}, \enquote{{MultiX}: Multi-modal {6G} experimental
  infrastructure}, \url{https://multix-6g.eu/}, 2025.

\bibitem{ref_Li2024}
Y.~Li, J.~Zhao, J.~Liao, and F.~Hu, \enquote{Cellular {V2X-Based} integrated
  sensing and communication system: Feasibility and performance analysis},
  \emph{Chin. J. Electron.}, vol.33, no.4, pp.1104--1116,
  \doi{10.23919/cje.2022.00.340}, 2024.

\bibitem{tianTVT}
Y.~Tian, J.~Wang, Y.~Wang, C.~Zhao, \emph{et~al.}, \enquote{Federated vehicular
  transformers and their federations: Privacy-preserving computing and
  cooperation for autonomous driving}, \emph{IEEE Trans. Veh. Technol.}, vol.7,
  no.3, pp.456--465, \doi{10.1109/TIV.2022.3197815}, 2022.

\bibitem{yuca2025survey}
N.~Yuca, F.~Klement, M.~Kroll, D.~R.~H. Spath, and R.~S.~T. Stuebiger,
  \emph{\enquote{A survey on privacy-preserving computing in the automotive
  domain}, arXiv e-Print}, arXiv:2508.01798, 2025.

\bibitem{ref_Xia2024}
J.~Xia, Y.~Liu, and L.~Tan, \enquote{Joint optimization of trajectory and task
  offloading for cellular-connected multi-{UAV} mobile edge computing},
  \emph{Chin. J. Electron.}, vol.33, no.3, pp.823--832,
  \doi{10.23919/cje.2022.00.159}, 2024.

\bibitem{Zhang2025}
H.~Zhang, B.~Li, J.~Huang, C.~Song, \emph{et~al.}, \enquote{A parallel
  multi-demonstrations generative adversarial imitation learning approach on
  {UAV} target tracking decision}, \emph{Chin. J. Electron.}, vol.34, no.4,
  pp.1185--1198, \doi{10.23919/cje.2024.00.082}, 2025.

\bibitem{itu2020radio}
\emph{Radio Regulations}, edition of 2020 ed., ITU, Geneva, Switzerland, ISBN
  978-92-61-31311-8, 2020.

\bibitem{kutynska2023legal}
A.~Kutynska and M.~Dei, \enquote{Legal regulation of the use of drones: Human
  rights and privacy challenges}, \emph{J. Int. Leg. Comm.}, vol.8, no.1,
  pp.39--55, \doi{10.32612/uw.27201643.2023.8.1.pp.39-55}, 2023.

\bibitem{yang2025generative}
Z.~Yang, G.~Chi, C.~Wu, H.~Liu, \emph{et~al.}, \emph{\enquote{Generative {AI}
  meets wireless sensing: Towards wireless foundation model}, arXiv e-Print},
  arXiv:2509.15258, 2025.

\bibitem{ref_MoE}
N.~Shazeer, A.~Mirhoseini, K.~Maziarz, A.~Davis, \emph{et~al.},
  \emph{\enquote{Outrageously large neural networks: The sparsely-gated
  mixture-of-experts layer}, arXiv e-Print}, arXiv:1701.06538, 2017.

\bibitem{ref_SHAP}
S.~M. Lundberg and S.-I. Lee, \enquote{A unified approach to interpreting model
  predictions}, \emph{Adv. Neural Inf. Process. Syst. (NeurIPS)}, vol.30, 2017.

\bibitem{ref_IG}
M.~Sundararajan, A.~Taly, and Q.~Yan, \enquote{Axiomatic attribution for deep
  networks}, in \emph{Proc. Int. Conf. Mach. Learn. (ICML)}, PMLR,
  pp.3319--3328, 2017.

\bibitem{ref_topk}
Y.~Lin, S.~Han, H.~Mao, Y.~Wang, and W.~J. Dally, \emph{\enquote{Deep gradient
  compression: Reducing the communication bandwidth for distributed training},
  arXiv e-Print}, arXiv:1712.01887, 2017.

\bibitem{ref_random_sparsification}
D.~Alistarh, T.~Hoefler, M.~Johansson, N.~Konstantinov, \emph{et~al.},
  \enquote{The convergence of sparsified gradient methods}, \emph{Adv. Neural
  Inf. Process. Syst. (NeurIPS)}, vol.31, 2018.

\bibitem{ref_joint_encoding_scheduling}
X.~Cao, Z.~Lyu, G.~Zhu, J.~Xu, \emph{et~al.}, \enquote{An overview on
  over-the-air federated edge learning}, \emph{IEEE Wirel. Commun.}, vol.31,
  no.3, pp.202--210, \doi{10.1109/MWC.005.2300016}, 2024.

\bibitem{Chen2024}
Y.~Chen, J.~Hu, J.~Zhao, and G.~Min, \enquote{{QoS}-aware computation
  offloading in {LEO} satellite edge computing for {IoT}}, \emph{Chin. J.
  Electron.}, vol.33, no.4, pp.875--885, \doi{10.23919/cje.2022.00.412}, 2024.

\bibitem{ref_FedAvgM}
T.-M.~H. Hsu, H.~Qi, and M.~Brown, \emph{\enquote{Measuring the effects of
  non-identical data distribution for federated visual classification}, arXiv
  e-Print}, arXiv:1909.06335, 2019.

\bibitem{ref_FedProx}
T.~Li, A.~K. Sahu, M.~Zaheer, M.~Sanjabi, \emph{et~al.}, \enquote{Federated
  optimization in heterogeneous networks}, \emph{Proc. Mach. Learn. Syst.
  (MLSys)}, vol.2, pp.429--450, 2020.

\bibitem{ref_yang2019federated}
Q.~Yang, Y.~Liu, Y.~Cheng, Y.~Kang, \emph{et~al.}, \enquote{Federated machine
  learning: Concept and applications}, \emph{ACM Trans. Intell. Syst.
  Technol.}, vol.10, no.2, pp.1--19, 2019.

\bibitem{ref_cao2020fltrust}
X.~Cao, M.~Fang, J.~Liu, and N.~Z. Gong, \emph{\enquote{Fltrust:
  Byzantine-robust federated learning via trust bootstrapping}, arXiv e-Print},
  arXiv:2012.13995, 2020.

\bibitem{MENG2025104085}
R.~Meng, B.~Xu, X.~Xu, M.~Sun, \emph{et~al.}, \enquote{A survey of machine
  learning-based physical-layer authentication in wireless communications},
  \emph{J. Netw. and Comput. Appl.}, vol.235, artilce no.104085, ISSN
  1084-8045, \doi{https://doi.org/10.1016/j.jnca.2024.104085}, 2025.

\bibitem{carlini2023certified}
N.~Carlini, F.~Tram{\`e}r, K.~Dvijotham, L.~Rice, \emph{et~al.},
  \enquote{(certified!!) adversarial robustness for free!}, in \emph{Proc. Int.
  Conf. Learn. Represent. (ICLR)}, 2023.

\bibitem{ref_DL}
I.~Goodfellow, Y.~Bengio, and A.~Courville, \emph{Deep learning}, MIT press
  Cambridge, vol.1, 2016.

\bibitem{lu2024heal}
Y.~Lu, Y.~Hu, Y.~Zhong, D.~Wang, \emph{et~al.}, \enquote{An extensible
  framework for open heterogeneous collaborative perception}, in \emph{Proc.
  Int. Conf. Learn. Represent. (ICLR)}, 2024.

\bibitem{lu2023coalign}
Y.~Lu, Q.~Li, B.~Liu, M.~Dianati, \emph{et~al.}, \enquote{Robust collaborative
  {3D} object detection in presence of pose errors}, in \emph{Proc. IEEE Int.
  Conf. Robot. Autom. (ICRA)}, pp.4812--4818, 2023.

\bibitem{chen2019fcooper}
Q.~Chen, X.~Ma, S.~Tang, J.~Guo, \emph{et~al.}, \enquote{{F-Cooper}: Feature
  based cooperative perception for autonomous vehicle edge computing system
  using {3D} point clouds}, in \emph{Proc. ACM/IEEE Symp. Edge Comput. (SEC)},
  pp.88--100, 2019.

\bibitem{xu2022v2xvit}
R.~Xu, H.~Xiang, Z.~Tu, X.~Xia, \emph{et~al.}, \enquote{{V2X-ViT}:
  Vehicle-to-everything cooperative perception with vision transformer}, in
  \emph{Proc. Eur. Conf. Comput. Vis. (ECCV)}, pp.107--124, 2022.

\bibitem{yin2021centerbased}
T.~Yin, X.~Zhou, and P.~Krahenbuhl, \enquote{Center-based {3D} object detection
  and tracking}, in \emph{Proc. IEEE/CVF Conf. Comput. Vis. Pattern Recognit.
  (CVPR)}, pp.11784--11793, 2021.

\bibitem{lang2019pointpillars}
A.~H. Lang, S.~Vora, H.~Caesar, L.~Zhou, \emph{et~al.},
  \enquote{{PointPillars}: Fast encoders for object detection from point
  clouds}, in \emph{Proc. IEEE/CVF Conf. Comput. Vis. Pattern Recognit.
  (CVPR)}, pp.12697--12705, 2019.

\bibitem{ref_ITU}
{International Telecommunication Union}, \enquote{{Report M.2410-0 (11/2017) -
  Minimum requirements related to technical performance for IMT-2020 radio
  interface(s)}}, \emph{Report}, No. M.2410-0, ITU-R, 2017,
  \urlprefix\url{https://www.itu.int/dms_pub/itu-r/opb/rep/R-REP-M.2410-2017-PDF-E.pdf}.

\bibitem{ref_TCR}
B.~Albert and T.~Tullis, \emph{Measuring the user experience: Collecting,
  analyzing, and presenting UX metrics}, Morgan Kaufmann, 2022.

\bibitem{Yue2024}
S.~Yue, Y.~Deng, G.~Wang, J.~Ren, and Y.~Zhang, \enquote{Federated offline
  reinforcement learning with proximal policy evaluation}, \emph{Chin. J.
  Electron.}, vol.33, no.6, pp.1360--1372, \doi{10.23919/cje.2023.00.288},
  2024.

\bibitem{ref_3GPP_ISAC}
{3GPP}, \enquote{Study on integrated sensing and communication},
  \emph{Technical Report}, No. 22.837, 3rd Generation Partnership Project,
  2022,
  \urlprefix\url{https://portal.3gpp.org/desktopmodules/Specifications/SpecificationDetails.aspx?specificationId=4044}.

\bibitem{ETSIISAC}
{ETSI}, \enquote{Integrated sensing and communications {(ISAC)}; use cases and
  deployment scenarios}, \emph{{Group Report}}, No. {ETSI GR ISC 001 V1.1.1},
  {ETSI}, 2025.

\bibitem{ref_3GPP_AI}
J.~Montojo, \enquote{{AI/ML for NR air interface}},
  \url{https://www.3gpp.org/technologies/ai-ml-nr}, 2022.

\bibitem{ldrdd2025}
S.~Buchanan, D.~Pai, P.~Wang, and Y.~Ma, \emph{Learning Deep Representations of
  Data Distributions}, Online,
  \url{https://ma-lab-berkeley.github.io/deep-representation-learning-book/}.,
  2025.

\end{thebibliography}

\end{document}